%% file: wd40pc.tex
\documentclass[fleqn,usenatbib]{mnras}

\usepackage{newtxtext,newtxmath}

\usepackage[T1]{fontenc}
\usepackage{ae,aecompl}
\usepackage{ragged2e}

\usepackage{graphicx}	
\graphicspath{{./figures/}}
\usepackage{amsmath}	
\usepackage{amssymb}	
\usepackage{booktabs}
\usepackage{soul}
\usepackage{lscape}

\newcommand{\gaia}{\textit{Gaia}}
\newcommand{\teff}{\mbox{$T_\mathrm{eff}$}}
\newcommand{\Msun}{\mbox{$\mathrm{M}_\odot$}}

\title[\gaia\ white dwarfs within 40\,pc II]{\gaia\ white dwarfs within 40\,pc II: the volume-limited northern hemisphere sample}

\author[J. McCleery et al.]{Jack McCleery,$^{1}$\thanks{E-mail: jack.mccleery@warwick.ac.uk (JM)}
Pier-Emmanuel Tremblay,$^{1}$
Nicola Pietro Gentile Fusillo,$^{2}$
\newauthor Mark A. Hollands,$^{1}$ Boris T. G{\"a}nsicke,$^{1}$ Paula~Izquierdo,$^{3,4}$ Silvia Toonen,$^{5}$ 
\newauthor Tim Cunningham,$^1$ and Alberto Rebassa-Mansergas$^{6,7}$
\\
$^{1}$Department of Physics, University of Warwick, CV4 7AL, Coventry, UK\\
$^{2}$European Southern Observatory, Karl-Schwarzschild-Str. 2, D 85748 Garching, Germany\\
$^{3}$Instituto de Astrof\'isica de Canarias, 38205 La Laguna, Tenerife, Spain\\
$^{4}$Departamento de Astrof\'isca, Universidad de La Laguna, 38206 La Laguna,Tenerife, Spain\\
$^{5}$Institute for Gravitational Wave Astronomy, School of Physics and Astronomy, University of Birmingham, Birmingham, \\B15 2TT, UK\\
$^{6}$ Departament de F\'{\i}sica, Universitat Polit\`{e}cnica de Catalunya, c/Esteve Terrades 5, 08860 Castelldefels, Spain\\
$^{7}$ Institut d'Estudis Espacials de Catalunya, Ed. Nexus-201, c/Gran Capit\`a 2-4, 08034 Barcelona, Spain\\
}

\date{Accepted XXX. Received YYY; in original form ZZZ}

\pubyear{2020}

\begin{document}
\label{firstpage}
\pagerange{\pageref{firstpage}--\pageref{lastpage}}
\maketitle

\begin{abstract}
We present an overview of the sample of northern hemisphere white dwarfs within 40\,pc of the Sun detected from \gaia\ Data Release 2 (DR2). We find that 521 sources are spectroscopically confirmed degenerate stars, 111 of which were first identified as white dwarf candidates from \gaia\ DR2 and followed-up recently with the William Herschel Telescope and Gran Telescopio Canarias. Three additional white dwarf candidates remain spectroscopically unobserved and six unresolved binaries are known to include a white dwarf but were not in our initial selection in the \gaia\ DR2 Hertzsprung-Russell diagram (HRD). Atmospheric parameters are calculated from \gaia\ and Pan-STARRS photometry for all objects in the sample, confirming most of the trends previously observed in the much smaller 20\,pc sample. Local white dwarfs are overwhelmingly consistent with Galactic disc kinematics, with only four halo candidates. We find that DAZ white dwarfs are significantly less massive than the overall DA population ($\overline{M}_\mathrm{DAZ} =$ 0.59\,\Msun, $\overline{M}_\mathrm{DA} =$ 0.66\,\Msun). It may suggest that planet formation is less efficient at higher mass stars, producing more massive white dwarfs. We detect a sequence of crystallised white dwarfs in the mass range from 0.6 $\lesssim M/\Msun\ \lesssim$ 1.0 and find that the vast majority of objects on the sequence have standard kinematic properties that correspond to the average of the sample, suggesting that their nature can be explained by crystallisation alone. We also detect 26 double degenerates and white dwarf components in 56 wide binary systems.

\end{abstract}



\section{Introduction}
Stars are born in groups with initial spatial and kinematic homogeneity but large-scale galactic dynamical disturbances due to spiral arms and mergers greatly affect their present day orbits. The precise astrometric and photometric observations from the \textit{Gaia} spacecraft \citep{gaiadr2} have improved our understanding of the formation and chemical evolution of stars in the Milky Way \citep[see, e.g.][]{haywood2018} and have recognised the role played by a merger with at least one satellite galaxy in the formation of the thick disc \citep{helmi2018}. As these stars age, deplete their nuclear energy source and evolve as giants and white dwarfs, their luminosity can change by a factor of up to $10^{8}$. It has therefore been a challenge to assemble representative stellar samples of all ages and masses. \textit{Gaia} DR2 has led to a major increase in the size of the local volume-limited sample, detecting the vast majority of main-sequence stars and white dwarfs within $\approx$ 100\,pc \citep{gaiaHR,gentile2018gaia}. This constitutes a unique snapshot of stars that have formed at all look-back times and with initial masses from the hydrogen burning limit to 10 \Msun. The detailed formation history of this present-day local stellar sample can help to constrain the overall formation history and radial migration of stars and their planets in the Milky Way \citep{minchev2013,fantin2019}.

While the advances made by \textit{Gaia} DR2 have been transformative, identifying up to 400\,000 sources within 100\,pc, a full understanding of the local stellar population is still a major challenge. The local \textit{Gaia} HRD is contaminated by distant sources with erroneous parallaxes as well as faint sources with improper sky background subtraction, with the recommended quality cuts reducing the size of the local sample by a significant factor \citep{gaiaHR}. Furthermore most of the local \textit{Gaia} sources do not have spectroscopic follow-up, with the spectral type completeness dropping to a small percentage beyond 20\,pc \citep[see, e.g.,][]{henry2018}. This largely prevents the determination of precise stellar parameters from the lack of atmospheric chemical abundances, magnetic field strengths and binary parameters. Furthermore, stellar modelling needs to be improved considering that currently employed stellar evolution tracks for white dwarfs and M dwarfs deviate from the empirical \textit{Gaia} HRD \citep{hollands2018gaia,parsons2018,morrell2019}. These modelling issues directly impact the characterisation of the bulk properties of nearby exoplanets around M dwarfs and evolved planetary systems at white dwarfs.

Surveys of nearby cool white dwarfs have historically used reduced proper motion as a proxy for distance coupled with targeted spectroscopic and astrometric follow-ups \citep{liebert1988,bergeron1997,limoges2015,subasavage2017}. Over the time, white dwarfs likely to be within 20-25\,pc were catalogued in a series of papers which highlighted the diversity of the local stellar remnant population and their space density \citep{holberg2002,holberg2008,sion2009white,giammichele2012know,holberg201625}. \textit{Gaia} DR2 has improved the completeness of these samples, recovering 130 known white dwarfs within 20\,pc and identifying 9 new candidates \citep{hollands2018gaia}. It has also allowed a fairly complete census of double degenerates or white dwarfs as part of a wide binary system. \citet{hollands2018gaia} have estimated the \textit{Gaia} DR2 detection probability to be close to 99 per cent of all white dwarfs at 20\,pc and it is not expected to change significantly for distances up to 70-100\,pc \citep{gentile2018gaia}. Previous studies have attempted to assemble larger volume-limited samples of spectroscopically confirmed white dwarfs, e.g. within 40\,pc in the northern hemisphere \citep{limoges2015} where the authors estimate the completeness at 65-80\%. This needs to be reviewed in light of new \textit{Gaia} DR2 catalogues of white dwarf candidates  \citep{gentile2018gaia,jimenez2018} and our recent spectroscopic follow-up of these new candidates in \citet[][hereafter Paper~I]{tremblay2020_WHT}. 

Advantages of using volume-complete samples of white dwarfs are plentiful: for deriving the local stellar formation history and the age of Galactic structures \citep{tremblay2014,limoges2015,isern2019,kilic2019,fantin2019}, studying binary evolution including mergers \citep{holberg2013,toonen2017,cheng2019,temmink2019}, statistics of evolved planetary systems \citep{zuckerman2010,hollands2018planet}, constraining the origin of stellar magnetism \citep{ferrario2015,landstreet2019}, deriving the initial-to-final-mass relation \citep{el-badry2018}, testing white dwarf model accuracy and spectral evolution \citep{tremblay2019,genest-beaulieu2019,coutu2019,ourique2019,blouin2019,gentile2020,cunningham2020}, and deepening our understanding on dense matter physics including crystallisation \citep{blouin2019,tremblayNature}. This work focuses on the spectroscopic volume-limited sample of white dwarfs within 40\,pc in the northern hemisphere. We use the \textit{Gaia} DR2 white dwarf catalogue of \citet{gentile2018gaia} as a starting point. The volume contained within the northern 40\,pc hemisphere is a factor of four larger than the all-sky 20\,pc sample of \citet{hollands2018gaia}, offering a significant advantage in terms of number statistics for the studies described above. 

Since April 2018 we have spectroscopically observed most of the new \textit{Gaia} white dwarf candidates within 40\,pc in the northern hemisphere. The spectral types and stellar parameters for more than one hundred new white dwarfs are presented in the companion Paper~I. In this work, we focus on the statistics of the overall 40\,pc sample, combining spectral types from the literature and Paper~I with a new photometric analysis using \textit{Gaia} and Pan-STARRS data. This approach is inspired by our earlier study of \textit{Gaia} DR2 white dwarfs within 20\,pc \citep{hollands2018gaia}. We describe the sample and its completeness in Section \ref{sec:sample}. We discuss the kinematics properties in Section \ref{sec:kin}, the sub-sample of binaries in Section \ref{sec:binary}, and discuss the space density, mass distributions, magnetism and crystallisation in Section \ref{sec:discussion}. We conclude in Section \ref{sec:conclusions}. 

\section{Sample}\label{sec:sample}
Our sample was obtained from a subset of the candidate white dwarf \gaia\ DR2 catalogue compiled by \citet{gentile2018gaia}. A simple cut was made to find all objects with parallaxes greater than 25 mas, i.e. $D< 40$\,pc, resulting in 1233 objects, 1048 of which are high probability white dwarfs ($P_\textrm{WD} > 0.75$). The parameter quantifying the probability of a source being a white dwarf was calculated in \citet{gentile2018gaia} by creating a distribution map in HRD space of both spectroscopically confirmed Sloan Digital Sky Survey (SDSS) white dwarfs and contaminants.

\begin{figure*}
	\includegraphics[width=\textwidth]{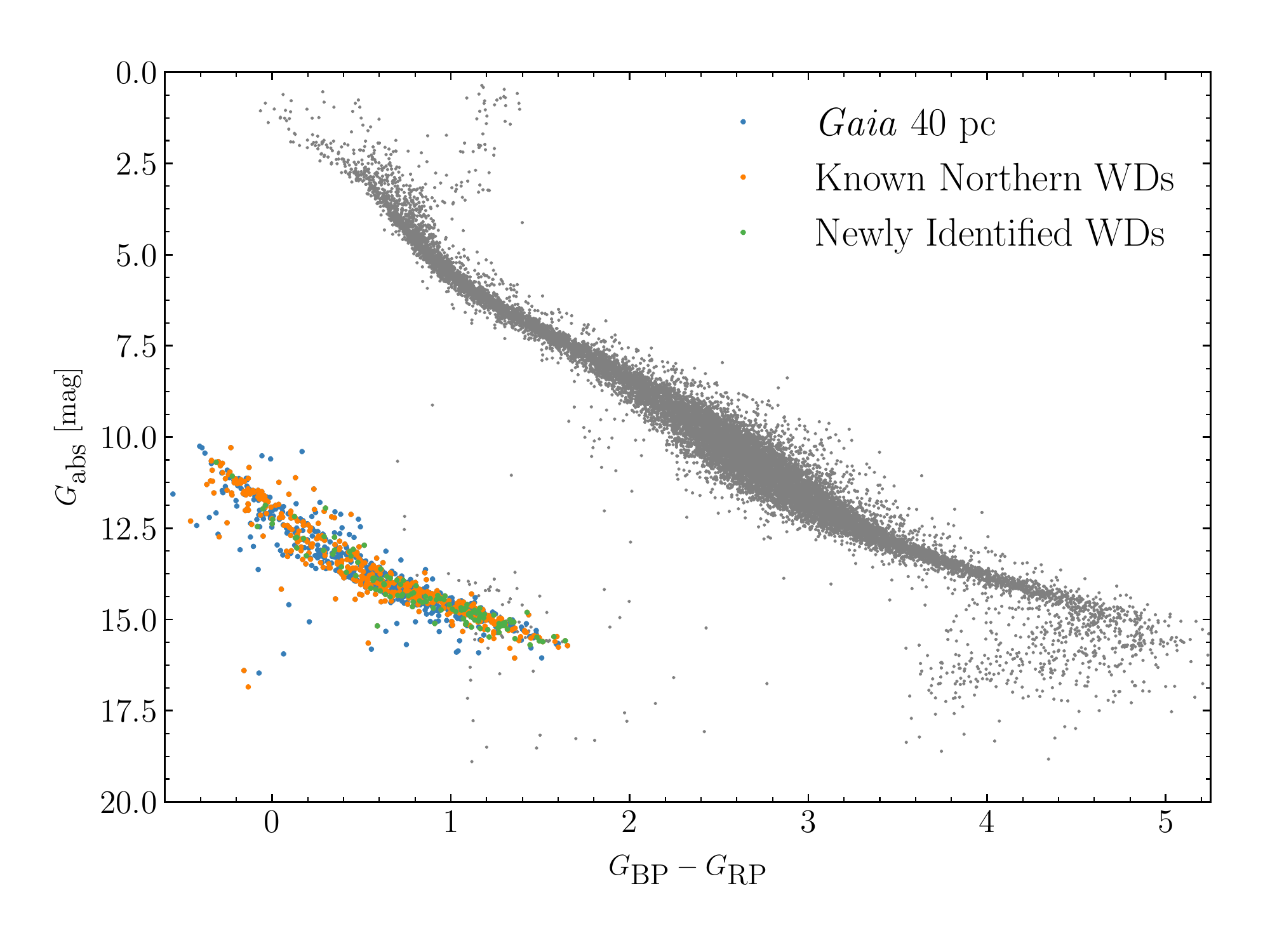}
    \caption{Hertzsprung-Russell diagram of the 40 pc population detected by \gaia. The high-probability \gaia\ 40\,pc white dwarf sample from \citet{gentile2018gaia}, both in the northern and southern hemispheres, is shown in blue, with the pre-\gaia\ northern white dwarfs (DEC $> 0$) shown in orange, and the newly confirmed white dwarfs from Paper~I in green. For reference, a \gaia\ 40\,pc sample is shown in grey, cleaned using the same methodology as \citet{gaiaHR}}.
    \label{fig:HR}
\end{figure*}

Of the 1233 candidates, 587 are located in the northern hemisphere. Cross-matching with catalogues of confirmed white dwarfs from the literature (e.g., \citealt{limoges2015}, \citealt{gianninas2011}, \citealt{kawka2012}, \citealt{subasavage2017}, full list in Table~\ref{tab:A1}), we find that 410 of the \textit{Gaia} sources correspond to white dwarfs with spectral types known before \textit{Gaia} DR2. Of the remaining 177 sources that were newly identified in \textit{Gaia}, 137 were observed spectroscopically and recently classified either in Paper~I\footnote{Paper~I also includes updated spectral types for six northern white dwarfs.} or in other recent papers \citep[see, e.g.,][]{landstreet2019,landstreet2020}. 111 of them turned out to be new white dwarfs while 26 are main-sequence stars or spurious \textit{Gaia} sources (see Paper~I). These 26 objects and a further 37 unobserved low probability white dwarf candidates are listed in the online material (Table~\ref{tab:A2})  and discussed in Section \ref{sec:non_wds}. As a consequence only three high probability white dwarf candidates specifically discussed in Section \ref{sec:missing} do not currently have spectral types. Our final volume-limited \textit{Gaia} sample of 521 confirmed white dwarfs and three high-probability white dwarf candidates is presented in the online material (Table~\ref{tab:A1}), with the description of the data content in Table \ref{tab:sample}. The objects are sorted by their WD\,J names as introduced in \citet{gentile2018gaia} while column 2 uses the WD name designation for objects cross-matched with the literature and known prior to \gaia\ DR2. Columns 3 to 11 repeat key data from \textit{Gaia} DR2 and \citet{gentile2018gaia}, while columns 12 to 23 report information on the atmosphere and stellar parameters as discussed in Section 2.1. Fig.~\ref{fig:HR} presents an overview in the HRD of the known and new white dwarfs within 40\,pc. 

\input{tables/headers.tex}
\input{tables/SpecTypes.tex}

We are missing local white dwarfs that were not selected by \citet{gentile2018gaia}. This can be  because the \textit{Gaia} DR2 five-parameter astrometric solution or colours are omitted or unreliable, or because the white dwarf is in an unresolved binary system and lies outside of their selection in the HRD. We have scanned the Montreal White Dwarf Database \citep{dufourMWDD} for objects within 40\,pc that are not in the full catalogue of \citet{gentile2018gaia}, finding a total of 22 white dwarfs presented in Table~\ref{tab:A4} that we now discuss in turn.

There are 13 white dwarfs for which a parallax value from \textit{Gaia} or other sources \citep{vanAltena95,vanLeeuwen07,leggett2018} confirms 40\,pc membership but that were not selected in \citet{gentile2018gaia} owing to missing, incomplete or unreliable \textit{Gaia} data. More than half of these objects are within 20\,pc and already discussed in \citet{hollands2018gaia}. In most cases these missing white dwarfs are nearby ($<$ 10\,pc), have large proper motions or are close to a bright stellar companion.

There are a further six multiple stellar systems which are known to include a white dwarf but are missing from our initial sample; UZ~Sex, V~EG~UMa, tet~Hya, V~DE~CVn, HD\,169889, and LHS 1817. These systems listed in Table~\ref{tab:A4} are all unresolved main-sequence + white dwarf binaries with complete \textit{Gaia} DR2 data. They were not selected by  \citet{gentile2018gaia} because the main-sequence star largely dominates the \gaia\ photmetry. As a consequence, these white dwarfs lay outside the region where they made their cuts on the HRD. As these binary systems were gathered from various literature sources, it remains a challenge to quantify their completeness within 40\,pc. We discuss these objects further in Section \ref{binaryWD+MS}. We also note a small number of cool and red objects ($G_{\rm BP}-G_{\rm RP} >$ 1.0) lying just above the white dwarf sequence in Fig.~\ref{fig:HR}, which covers both the southern and northern hemispheres. These are unobserved low probability white dwarf candidates from \citet{gentile2018gaia} and are listed in Table\,\ref{tab:A2}, a small number of which may be double degenerate binaries missing from our final sample.

Finally, Table~\ref{tab:A4} includes three white dwarfs with no parallax from any source, but that are possible 40\,pc members based on previously published spectroscopic or estimated photometric distances. We do not include any of the missing white dwarfs listed in Table~\ref{tab:A4} in the following statistical analysis to ensure homogeneity of the data. We conclude that our 40\,pc sample of Table~\ref{tab:A1} is at most 96 per cent complete, but that the final completeness is very likely to be close to that value \citep{gentile2018gaia}.

We consider 33 additional northern objects from the catalogue of \citet{gentile2018gaia} that are within 2$\sigma$ of a parallax of 25 mas separately in Table~\ref{tab:A3}. For simplicity the table merges together 17 confirmed white dwarfs (nine of which were observed in Paper~I), two main-sequence contaminants and 14 white dwarf candidates that have not been observed, only three of which are high probability white dwarf candidates. A few of these white dwarfs may turn out to be 40\,pc members with the improved astrometry from \gaia\ DR3, but to ensure that our statistics represent a volume-limited sample, we do not include any of these additional objects in our main analysis below.

\subsection{Spectral types and atmospheric parameters}

For each 40\,pc white dwarf we have gathered the spectral type from the literature, with the references given in Table~\ref{tab:A1}. A breakdown of the spectral types can be seen in Table \ref{tab:spect_table} (see, e.g., \citealt{sion83} for spectral type definitions). In the majority of cases, the spectral type is sufficient to conclude that the dominant constituent of the atmosphere is either hydrogen or helium, e.g. from the presence or absence of hydrogen Balmer lines. We do not have evidence of any carbon-dominated atmosphere white dwarf in the northern 40\,pc sample, although three magnetic white dwarfs with unknown absorption bands (spectral type DX or DXP) have an ambiguous atmospheric composition. Below effective temperatures ($T_{\rm eff}$) of $5000$\,K, the vast majority of white dwarfs are of featureless DC spectral type and it is not straightforward to determine the atmospheric composition \citep{blouin2019}. A few of these cool objects are of metal-rich DZ or DZA spectral types, allowing to constrain the atmospheric composition based on the effect of helium or hydrogen broadening on the metal lines. 

The classification into different spectral subtypes is sensitive to the signal-to-noise ratio (S/N) of the spectroscopic observations. In particular, the detection of weak metal or carbon lines as well as weak magnetic fields ($B \lessapprox$ 500\,kG) is only possible at sufficiently high S/N. This is also the case for cool hydrogen atmosphere DA white dwarfs with only weak H$\alpha$ lines at $T_{\rm eff} \approx 5000$\,K. A visual inspection of the published spectroscopic studies of local white dwarfs \citep{giammichele2012know,limoges2015,gentile2018gaia,tremblay2020_WHT} indeed reveals variations in S/N. Furthermore, when available we have updated our spectral types based on dedicated high-resolution observations or spectropolarimetry, which have observed metal lines or magnetic fields not seen in lower resolution observations. Many of these high-resolution surveys have favoured close (20\,pc) and brighter white dwarfs \citep[see, e.g.,][]{zuckerman2003,landstreet2019}.

\begin{figure}
	\includegraphics[width=\columnwidth]{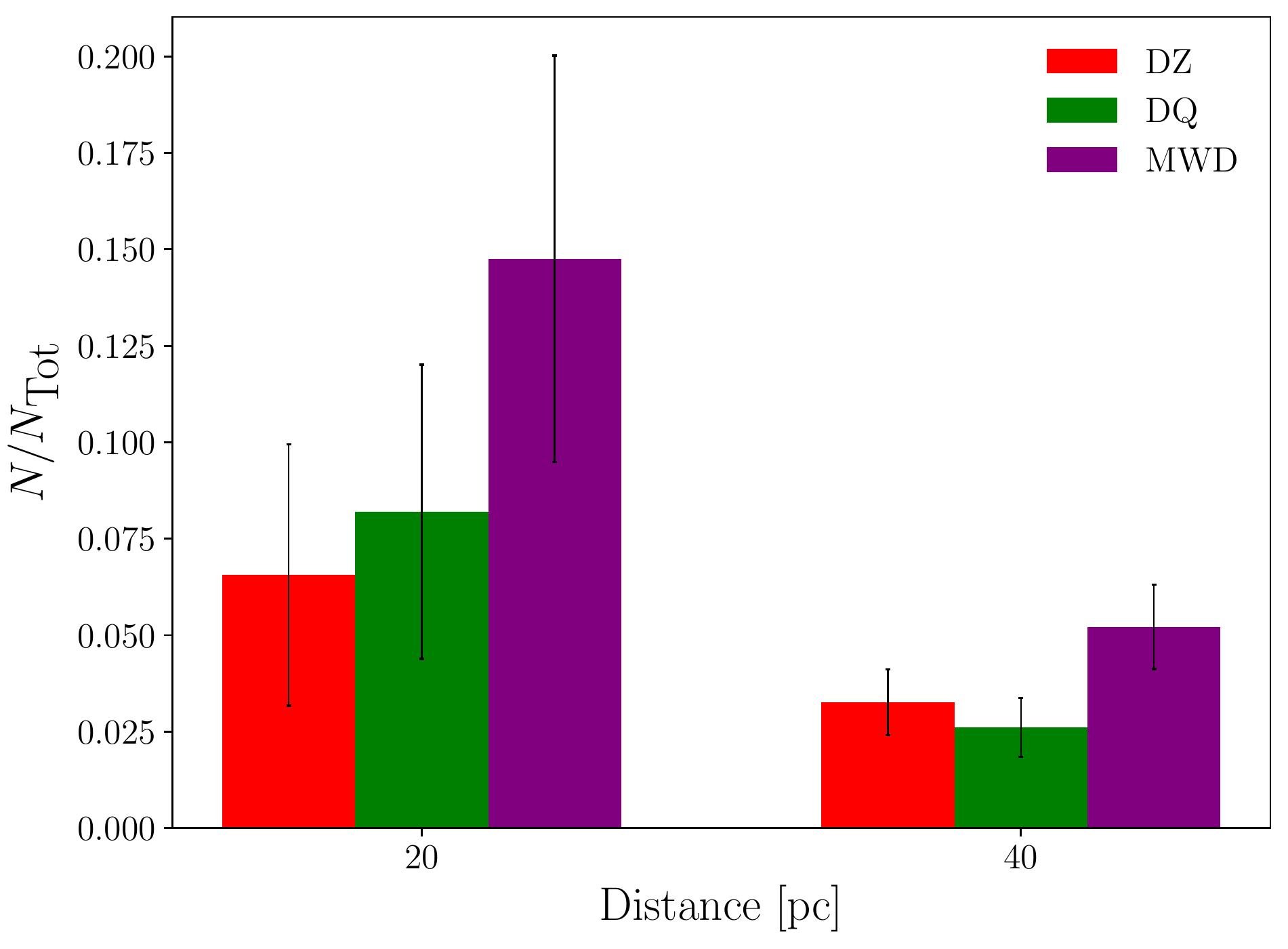}
    \caption{Distribution of spectral types as a function of distance with DZs shown in red, DQs in green, and magnetic white dwarfs in purple. The first bins correspond to the 20\,pc sample and the second bins to the 20\,pc $< d <$ 40\,pc sample.}
    \label{fig:spt_dist}
\end{figure}

The distribution of spectral types as a function of distance is shown in Fig.~\ref{fig:spt_dist}. The fractions of magnetic, DZ, and DQ white dwarfs within 20\,pc $< d <$ 40\,pc are all within 2$\sigma$ of those found for the 20\,pc sample, implying that biases due to S/N of the observations are only marginally significant given the small size of the 20\,pc sample. This nevertheless suggests that deeper observations of the 40\,pc sample could lead to an increase of a factor of about two in the number of detected subtypes, and one should be cautious in the determination of the absolute fraction of magnetic, DZ, and DQ white dwarfs using the 40\,pc sample.

While atmospheric parameters derived from fits to optical spectroscopy can be gathered in the literature for the warmest objects in the sample, here we take advantage of the high-precision broadband photometry and astrometry from \textit{Gaia} to derive a homogeneous set of high-precision atmospheric parameters. It should be noted that a systematic offset at the few percent level has been observed between the \textit{Gaia} and spectroscopic temperature scales, both for warm DA or DB white dwarfs as well as cool DA stars \citep{tremblay2019,tremblay2020_WHT,genest-beaulieu2019}, hence one should be cautious about the absolute temperature and mass scales. 

We used pure-hydrogen \citep{hydrogenmodelpier2011}, pure-helium \citep{heliummodelbergeron2011}, and mixed model atmospheres \citep{mixedmodelpier2014} to calculate $T_{\rm eff}$ and surface gravity ($\log g$) based on \textit{Gaia} $G$, $G_{\rm BP}$ and $G_{\rm RP}$ photometry as well as parallax for each white dwarf in the sample. The fitting method is the same as that described in \citet{gentile2018gaia}. In brief, we rely on the Ly$\alpha$ broadening of \citet{kowalski2006} and the mass-radius relation of \citet{fontaine2001} for thick (H-atmospheres) or thin (He-atmospheres) hydrogen envelopes and C/O-cores. The only differences are that we have completely neglected reddening, which is justified for the 40\,pc sample, and we have added the option of mixed He/H atmospheres. 

Given that we have now secured a breakdown of the spectral types for all but three objects in the sample, we adopt the atmospheric parameters based on the inferred composition. This is an improvement with respect to \citet{gentile2018gaia} who provided both H- and He-atmosphere options. Assignations of the different spectral types to an atmospheric composition are given in Table \ref{tab:spect_table}. 

\begin{figure}
	\includegraphics[width=\columnwidth]{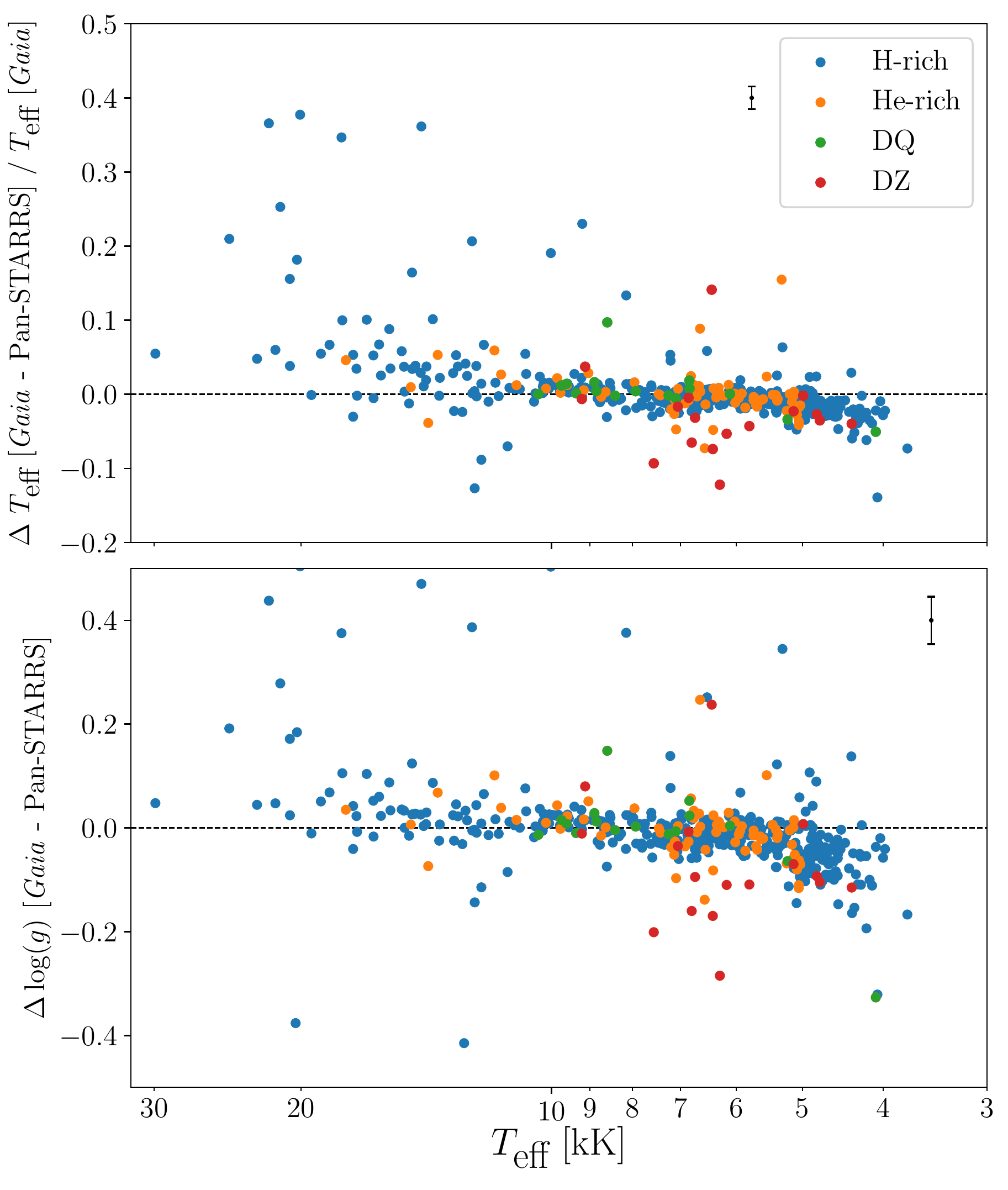}
    \caption{Difference between the measured \gaia\ and Pan-STARRS atmospheric parameters; fractional \teff\ difference is shown in the top panel and absolute $\log g$ difference in the bottom. Both are shown as a function of \teff. The average formal errors on the difference are shown in the top right of both panels.}
    \label{fig:delta_logg}
\end{figure}

We have taken advantage of the essentially complete northern hemisphere coverage of the Pan-STARRS survey \citep{panstarrs} to derive an independent set of atmospheric parameters based on $grizy$ photometry and \textit{Gaia} parallaxes. It was demonstrated in \citet{gentile2018gaia} that Pan-STARRS and \textit{Gaia} photometry are in good agreement for warm DA white dwarfs, and here we explore this further with the cool white dwarfs within 40\,pc. Fig.~\ref{fig:delta_logg} demonstrates that $\log g$ and $T_{\rm eff}$ values are generally within agreement at the few percent level and within combined error bars. Most of the outliers are in crowded areas of the sky or close to a bright companion (see Paper~I). Below $T_{\rm eff} \approx 5000$\,K, \textit{Gaia} colours are systematically redder than Pan-STARRS colours (and predicted white dwarf cooling sequences, see, e.g., \citealt{bergeron2019}), resulting in systematically lower temperatures and surface gravities. It is unclear if this is a calibration effect or inaccurate physics influencing the \textit{Gaia} and Pan-STARRS bandpasses differently. 

For most local white dwarfs, the \textit{Gaia} passbands are broad enough that the colours are not significantly impacted by average metal pollution or magnetic fields. Furthermore, the large majority of the objects are cool enough that metal line UV blanketing is expected to be negligible. Fig.~\ref{fig:delta_logg} demonstrates that we obtain similar atmospheric parameters with the much narrower Pan-STARRS filters, apart from DZ white dwarfs which appear to be more scattered. Overall, we conclude that using pure-H, pure-He and mixed H/He models is not a major concern for our analysis. 

There are two cool DQ and DQpec with very broad and deep absorption bands. There are also four ultra-cool DC white dwarfs with significant collision-induced absorption (CIA). Those do not have meaningful \textit{Gaia} or Pan-STARRS atmospheric parameters and are flagged with a He+CIA composition in Table~\ref{tab:A1} where appropriate. 

\citet{gentile2018gaia} and \citet{bergeron2019} have shown that cool helium-rich white dwarfs with 7000\,K $\lesssim T_{\rm eff} \lesssim$ 11\,000\,K have larger masses than their warmer DB/DBA progenitors when fitted with pure helium models. This is inconsistent with an expected stellar evolution at constant mass. \citet{bergeron2019} have demonstrated that by adding trace hydrogen in the helium-rich atmospheres, the predicted masses are lower and in much better agreement with the expectation of evolution models. The presence of hydrogen is also consistent with DC and DZ white dwarfs being the cooler counterparts of DB/DBA white dwarfs, for which of the order of 50 per cent of objects have detectable hydrogen (H/He > $10^{-7}$ in number) with a median value of H/He $\approx$ $10^{-5}$ \citep{rolland2018}. Such trace hydrogen abundance is not detectable in helium-rich DC or DZ white dwarfs cooler than about 11\,000\,K for typical low- or medium-resolution observations \citep{rolland2018} but is needed for the large majority of these cool objects to fit the predicted cooling tracks. As a consequence, we have adopted helium-rich model atmospheres with H/He = $10^{-5}$ for all helium-rich spectral types warmer than 7000\,K. 

Below $T_{\rm eff}$ = 7000\,K, our mixed model atmospheres \citep{mixedmodelpier2014} predict a blue hook in \textit{Gaia} colours due to CIA opacities, which is not observed for the vast majority of white dwarfs \citep{gentile2020}. This blue hook is also not predicted in more recent mixed H/He model atmospheres discussed in \citet{blouin2019}. As a consequence, we use instead pure-helium solutions for all helium-rich objects below 7000\,K, where the bifurcation to high masses is not observed. Finally, most objects below 5000\,K are DC white dwarfs and it is not possible to determine the atmospheric composition based on spectroscopy and optical photometry alone, and challenging even with near-IR photometry \citep{gentile2020}. Therefore, we assign a pure-hydrogen composition for all objects below that temperature. We note that in the range 7000\,K $\gtrsim T_{\rm eff} \gtrsim$ 4500\,K our pure-H and pure-He solutions differ only by a few percent. As a consequence our $T_{\rm eff}$ and $M$ estimates are, in principle, still robust even with an unconstrained composition.

\begin{figure}
	\includegraphics[width=\columnwidth]{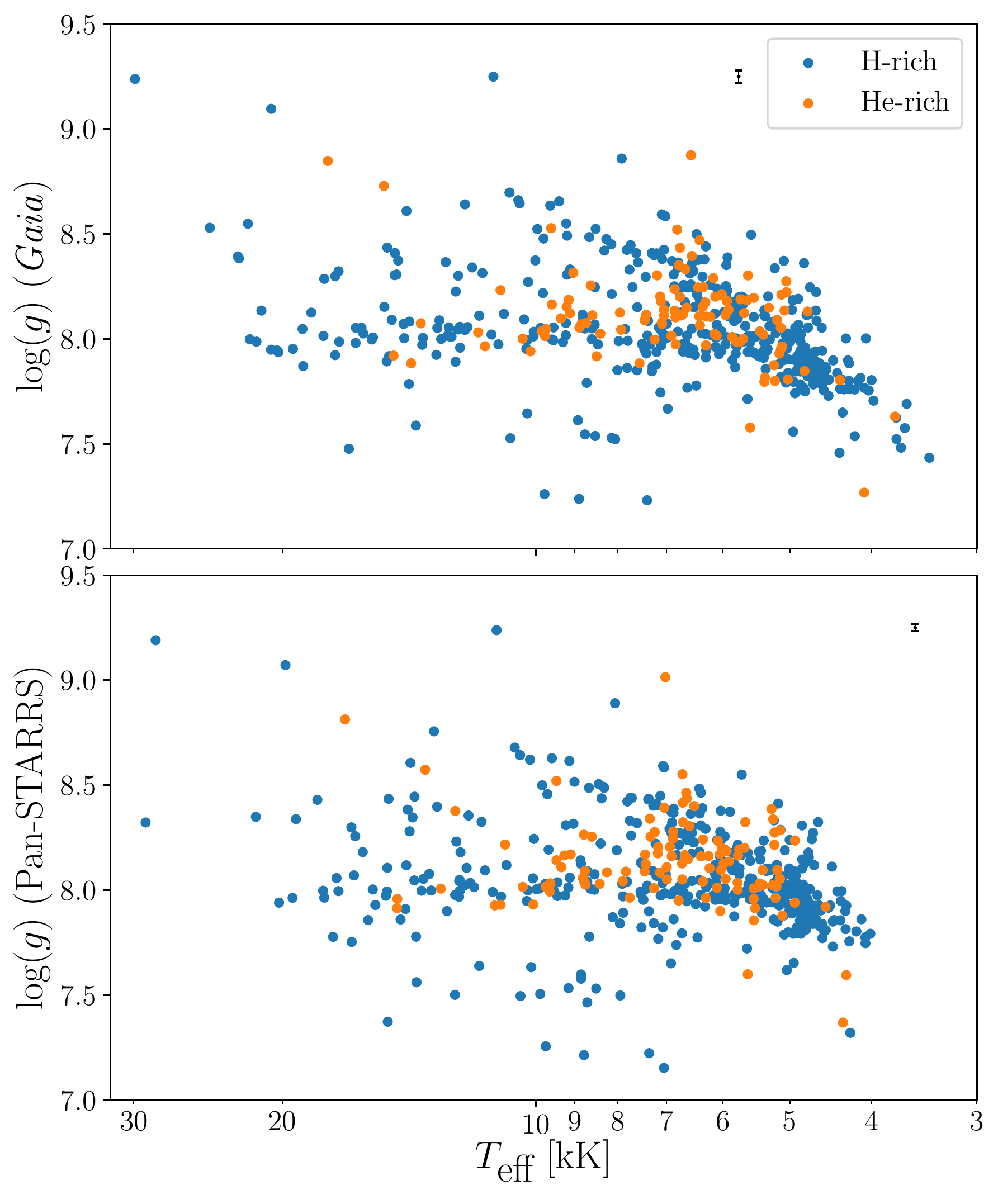}
    \caption{Distribution of $\log g$ vs \teff\ for \gaia\ DR2 (top panel) and Pan-STARRS (bottom) photometry. We rely on \gaia\ DR2 astrometry in both cases. H-rich objects are shown in blue, whilst He-rich are shown in orange. The average errors are shown in the top right of both panels. Reliable \textit{Gaia} and Pan-STARRS atmospheric parameters are available for 517 and 492 objects in the sample, respectively.}
    \label{fig:logg_v_teff}
\end{figure}

Fig.~\ref{fig:logg_v_teff} shows the $\log g$ versus $T_{\rm eff}$ distribution, using both \textit{Gaia} and Pan-STARRS data, for 40\,pc white dwarfs where six ultra-cool and DQ white dwarfs with unreliable \textit{Gaia} parameters are excluded. The crystallisation sequence \citep{tremblayNature} is clearly seen as a parabola with $\log g$ increasing as a function of $T_{\rm eff}$, starting at 6000\,K and $\log g \approx 8.1$ and increasing to $\log g \approx 8.7$ at 10\,000\,K. In the 100\,pc sample of \citet{tremblayNature} and 200\,pc sample of \citet{cheng2019} the crystallisation sequence can be seen to extend to higher temperatures and surface gravities, but given the limited volume of the 40\,pc sample, the $\log g > 8.7$ region (corresponding to $M \gtrsim 1.05$ M$_{\odot}$) is under-populated. Most white dwarfs on the crystallisation sequence are of DA spectral type, as highlighted in \citet{tremblayNature}. We discuss the crystallisation sequence further in Section \ref{sec:crystal}. 

We observe that parameters for H- and He-rich atmospheres have no obvious offset when using mixed He/H instead of pure-He models, as highlighted in \citet{bergeron2019} and effectively correcting the so-called \textit{bifurcation} problem \citep{gentile2018gaia}.

For $T_{\rm eff} \lesssim 5500$\,K, $\log g$ values decrease with decreasing temperature as previously highlighted for the 20\,pc sample in \citet{hollands2018gaia}. For fixed mass-radius relation, parallax and apparent magnitude, the effective temperature and surface gravity correlate because of the Stefan-Boltzmann law. This suggests that the problem could arise either from photometric colours that are predicted too blue or absolute magnitudes that are predicted too faint. A very similar pattern is seen with independent grids of models \citep{blouin2019}, and for any atmospheric composition including pure-H, pure-He or mixed \citep{bergeron2019}. The problem is clearly seen using either \textit{Gaia} or Pan-STARRS photometry, although the issue is slightly worse with \textit{Gaia} (see Fig.~\ref{fig:delta_logg}). It is not expected to be a real astrophysical effect as white dwarfs at these temperatures have cooling ages in the range $\approx$ 5 -- 10\,Gyr. Stellar population models predict a constant mean mass for white dwarfs with cooling ages smaller than about 10\,Gyr \citep{tremblay2016}. Therefore, the lower than average $\log g$ (or $T_{\rm eff}$) values are more likely to be explained from an issue with the model atmospheres. The problem appears for white dwarfs with a vast range of cooling ages, hence this should be taken into account properly in order to extract meaningful stellar formation histories when transforming white dwarf parameters to initial stellar parameters \citep{tremblay2014}.

\subsection{Missing spectral types}\label{sec:missing}

There are three objects for which ground-based spectroscopic observations are challenging and currently not available in the literature. All three are listed in Table~\ref{tab:A1} but without a spectral type. 

\textbf{J050600.41+590326.89} is a high proper-motion (346.95 mas/yr) {\it Gaia} source that was blended with a distant background star at the time of our observations in Paper~I. The object is extremely faint ($G$ = 19.65) and blue ($G_{\rm BP}-G_{\rm RP} = -0.1335$). Due to the faintness, the Gaia measurements have high astrometric excess noise, suggesting we must be careful about its status, either as a rare ultra-cool white dwarf or as a non-degenerate source. For this reason the atmospheric parameters are omitted in Table~\ref{tab:A1}.

\textbf{J055602.01+135446.71} is a high proper-motion (608.06 mas/yr) {\it Gaia} source that was close to a distant background star at the time of our observations in Paper~I. The white dwarf candidate is itself a wide companion to the bright star HD~39881 of spectral class G8 at 47.7$^{\prime\prime}$ separation. The presence of a wide companion at the same distance and proper motion makes the identification as a white dwarf fairly secure.

\textbf{J110143.04+172139.39} is in the glare (17$^{\prime\prime}$ separation) of the background F-type main-sequence star HD~95518 with $G= 8.37$. 

\subsection{Non-white dwarfs}\label{sec:non_wds}

Table~\ref{tab:A2} lists 26 objects confirmed as main-sequence stars and a further 37 unobserved low probability white dwarf candidates from the initial sample of \citet{gentile2018gaia}. None of the objects they selected within the northern 40\,pc sample were known as non-white dwarfs in the literature before \textit{Gaia}. All 26 main-sequence stars were therefore identified as part of our recent spectroscopic follow-up of white dwarf candidates in Paper~I. Seven out of the 26 \textit{Gaia} sources found to be main-sequence stars were high probability white dwarf candidates ($P_{\rm WD}>0.75$). This is a relatively small fraction of the final white dwarf sample (1.3 per cent), and therefore it largely confirms the cleanness of high-probability white dwarf samples selected from \citet{gentile2018gaia}. Paper~I concludes that a problem with astrometry is the most likely explanation for the spurious low luminosity of these stellar sources that are located well within the white dwarf cooling track according to the \textit{Gaia} DR2 HRD. In all but one case\footnote{In one case the \textit{Gaia} source does not correspond to a real object on the sky.} the Pan-STARRS colours do agree with \textit{Gaia} colours. 

The 40\,pc sample of \citet{gentile2018gaia} contained 66 low probability white dwarf candidates in the northern hemisphere. Among those, eight objects are white dwarfs that were known before \textit{Gaia}. A further 21 sources were recently observed in Paper~I but only 2 objects turned out to be white dwarfs. There are 37 objects that have not been spectroscopically observed, and among them there could be a few white dwarfs. We have attempted to use kinematics or infrared colours to reveal their nature but found that any cut on only one of these quantities would also eliminate known white dwarfs. 
\textit{Gaia} DR3 is expected to help in defining a cleaner and more complete 40\,pc sample.

\section{Kinematics}\label{sec:kin}

\begin{figure}
	\includegraphics[width=\columnwidth]{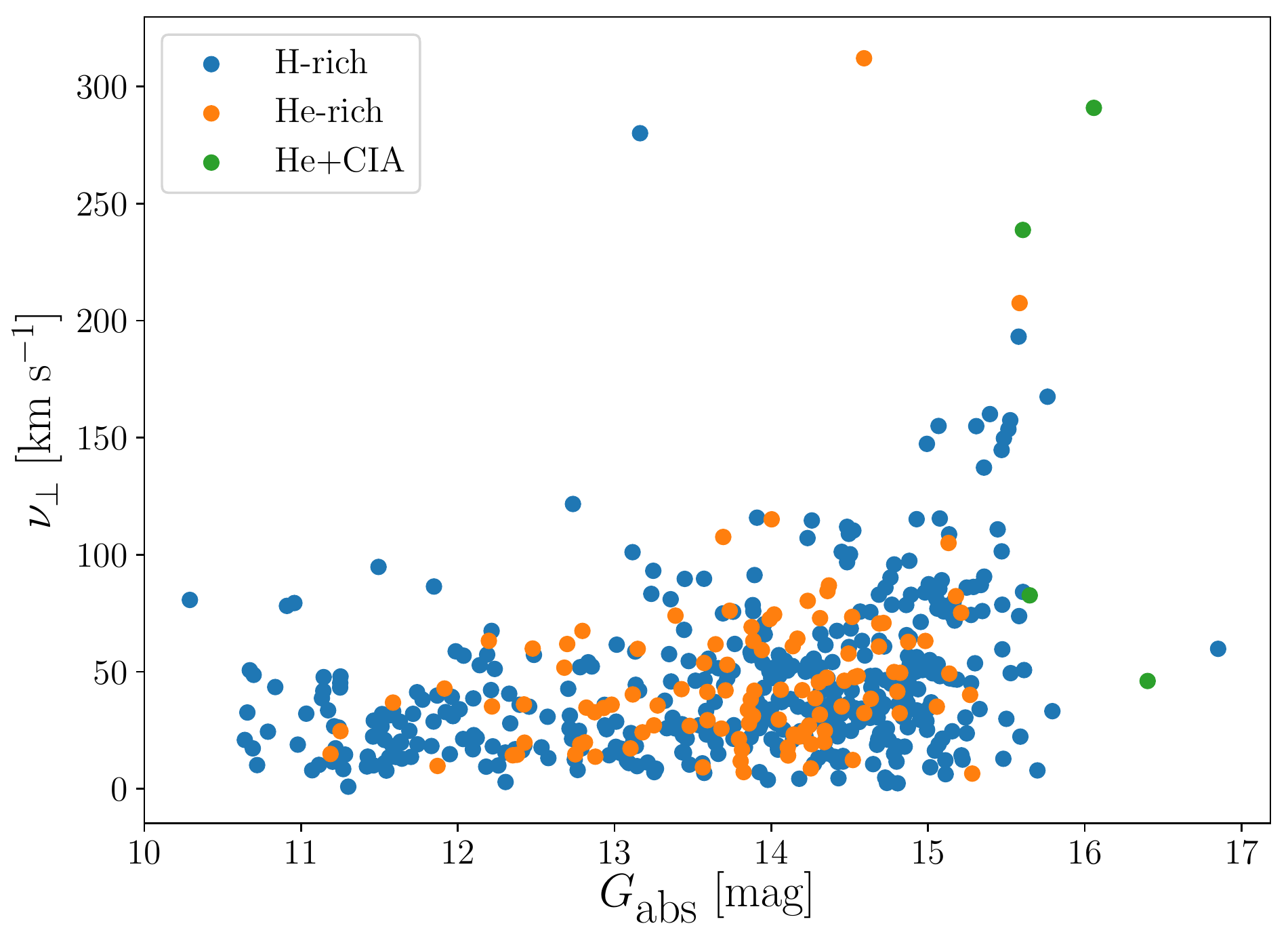}
    \caption{Tangential velocities of the sample as a function of $G_\textrm{abs}$ magnitude. H-rich objects are shown in blue, whilst He-rich objects are shown in orange. Four ultracool white dwarfs are shown in green.}
    \label{fig:tanv}
\end{figure}

We can calculate tangential velocities, $\nu_{\perp}$, for our sample with \gaia\ data alone using
\begin{equation}
    \nu_\perp = 4.7405\sqrt{\mu_{RA}^2 + \mu_{Dec}^2} / \varpi
	\label{eq:tanv}
\end{equation}
where $\mu_{Ra/Dec}$ are the right ascension/declination components of the proper motion in mas yr$^{-1}$, and $\varpi$ is the parallax of the white dwarf in mas. Fig.~\ref{fig:tanv} shows $\nu_{\perp}$ as a function of $G_\textrm{abs}$, which is a proxy for cooling age. We expect that throughout their full evolution on the main-sequence and white dwarf cooling sequence, these stars will be subject to kinematic heating by the Galactic potential, with increasing $\nu_{\perp}$ as a function of total age \citep{seabroke2007}. While white dwarf cooling age is only a fraction of the total age, larger cooling ages lead to, on average, larger total ages, and therefore cooling age should also correlate with kinematics, as observed in Fig.~\ref{fig:tanv}. In a future study we will consider total ages, which depend more critically on the absolute precision of our white dwarf masses. 

Fig.~\ref{fig:tanv} also reveals that the vast majority of 40\,pc white dwarfs are consistent with Galactic disc kinematics. Only four objects are clear outliers in the diagram:

\textbf{J222547.07+635727.37}, $\nu_\perp =\ 312$ km s$^{-1}$, is a white dwarf of spectral type DC discovered by \gaia\ \citep{tremblay2020_WHT} and previously identified as a halo white dwarf candidate in \citet{kilic2019}. Our low surface gravity ($\log g$ = 7.80 $\pm$ 0.03) suggests a large total age in agreement with halo membership.

\textbf{J110217.52+411321.18}, $\nu_\perp =\ 291$ km s$^{-1}$, is an ultracool DC white dwarf first discovered by SDSS and was suggested to be an old halo object with a cooling age of 11 Gyr by \citet{Hall2008halowd}. Because of the strong CIA absorption in the optical, it is not possible to constrain the stellar mass with \textit{Gaia} or Pan-STARRS data alone. 

\textbf{J174950.15+824626.06}, $\nu_\perp =\ 280$ km s$^{-1}$, is a known 20\,pc DA white dwarf (WD 1756+827) first suggested to be halo candidate by \citet{fuchs199820pchalowd} and discussed further in \citet{hollands2018gaia} and \citet{kilic2019}.

\textbf{J034646.52+245602.67}, $\nu_\perp =\ 239$ km s$^{-1}$, is another ultracool DC white dwarf which was previously known for its high proper-motion and halo membership (WD 0343+247; \citealt{hambly1997halowd,kilic2019}).

\section{Binarity}\label{sec:binary}

\subsection{Wide binaries}\label{sec:wide-binary}

To search for wide binaries within our sample, both white dwarf + main-sequence (WD+MS) and double white dwarfs (WD+WD), we employ the same technique as \citet{hollands2018gaia}. In brief, we perform a cylindrical search around each white dwarf in our sample to a projected separation ($D_\perp$) of 1\,pc. For each object found, a cut was made in the absolute difference in radial distances ($\Delta D_\parallel$), also at 1\,pc. In addition we have used the standard quality cuts of \citet{gaiaHR} on astrometric excess noise and colour excess on all wide stellar companions. The remaining stars within the search volume were then checked for consistent tangential velocities ($\Delta\nu_\perp$). Those with large separations and large tangential velocity differences that correspond to physically unbound systems for a total mass of 2 M$_{\odot}$ on circular orbits were rejected as being non-companions.  

\begin{figure}
	\includegraphics[width=\columnwidth]{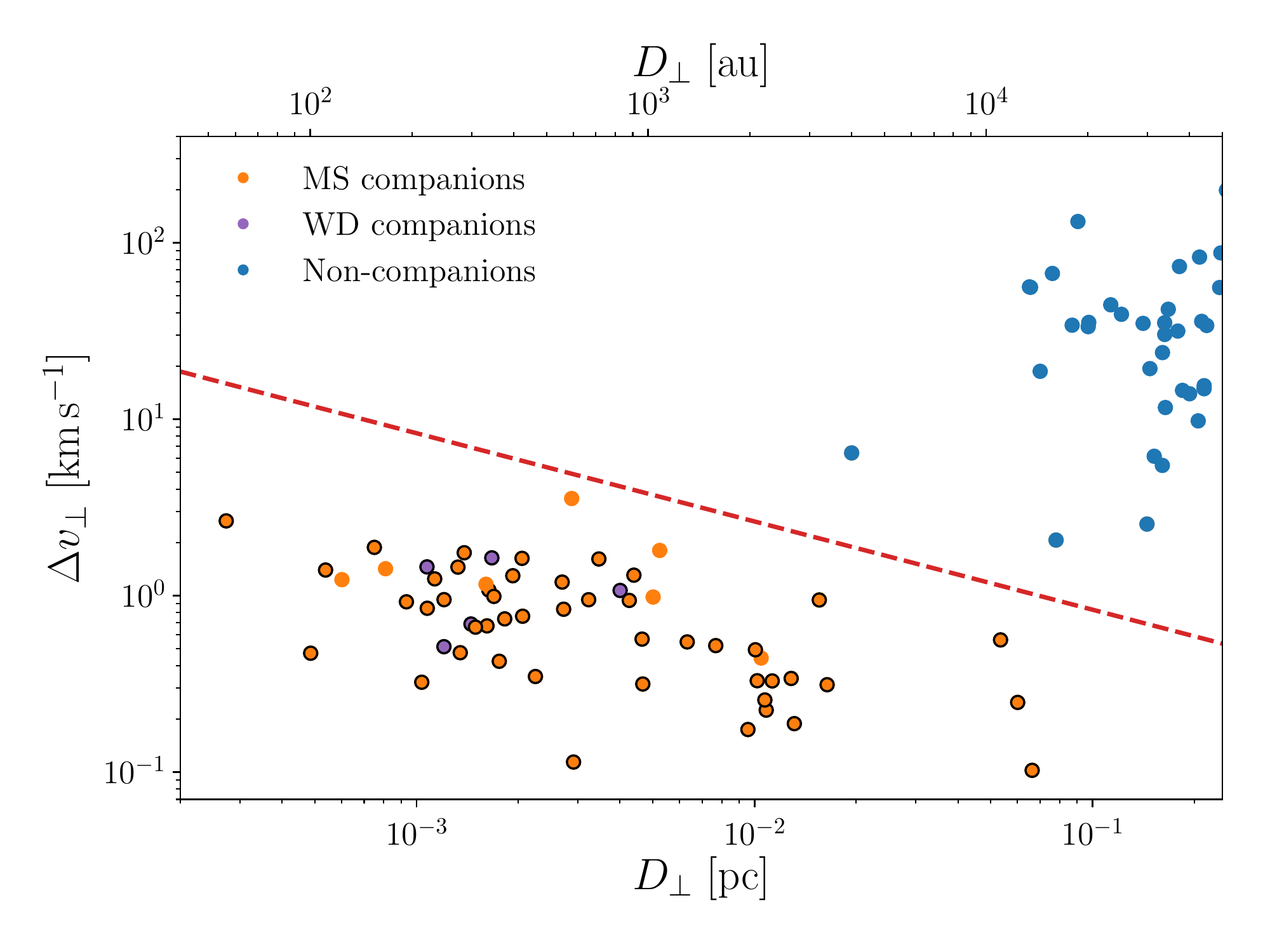}
    \caption{Tangential velocity differences as a function of projected separation for all \gaia\ sources within 50\,000 AU of the white dwarfs in our sample. Companions also found in \citet{el-badry2018binary} are outlined in black. The red dashed line indicates the maximum tangential velocity difference for a system of two 1\,\Msun\ stars on circular orbits with a semi-major axis of $D_\perp/2$. Our final candidates are shown in orange (WD+MS) and purple (WD+WD). Blue candidates are rejected because of their nonphysical separation in position-velocity space.}
    \label{fig:wide_binaries}
\end{figure}

We find a total of 56 binary system candidates in our sample, these are shown in Fig.~\ref{fig:wide_binaries} and listed in Table~\ref{tab:A5}. Those consist of 47 WD+MS, 1 WD+MS+MS and 8 WD+WD binaries. From a similar \textit{Gaia} DR2 selection, \citet{el-badry2018binary} found 43 WD+MS and 8 WD+WD within the northern 40\,pc sample, all of which we recover. \citet{el-badry2018binary} neglected triple systems and used slightly stricter selection rules as they were aiming at minimising contamination rather than maximising completeness in their much larger volume of 200\,pc.

In comparison, \citet{hollands2018gaia} found 23 wide binary systems within 20\,pc. Of these, 21 are WD+MS binaries, and only two are WD+WD binaries. Extrapolating the space density from the 20\,pc sample we could expect 92 $\pm$ 19 systems within 40\,pc. However, it is found in Section~\ref{sec:discussion} that the space density itself is seen to decrease with distance for all white dwarfs (2--5 per cent effect). Furthermore, \textit{Gaia} resolving power is decreasing with distance and therefore 8--13 per cent lower numbers are expected for the 40\,pc sample \citep{toonen2017,hollands2018gaia}. Given these two effects and the relatively large error bars from number statistics, the difference in space density of wide systems including a white dwarf between the 20 and 40\,pc samples is only marginally significant.

Based on binary population synthesis for the 50\,pc \gaia\ sample and re-scaled to our volume, \citet{toonen2017} predict 85 -- 150 resolved WD+MS and 60 -- 112 resolved WD+WD systems. These numbers can decrease by 15--30 per cent if we consider the possible disruption of weakly bound binaries by Galactic interactions and stellar winds. The number of observed WD+MS systems is low but marginally in agreement with model predictions if a large fraction of initially formed WD+MS systems were disrupted. The issue of missing WD+WD systems is already discussed at length in \citet{toonen2017}, although the deficit is now strongly confirmed by the 40\,pc sample.

\citet{el-badry2018binary} discuss the properties of wide binaries including a white dwarf from the much larger 200\,pc sample. They find that their orbital separation distribution differs from that of wide binaries including two stars, suggesting that velocity kicks from mass loss during stellar evolution play an important role in these systems. \citet{el-badry2018binary} notice that the deficit is enhanced for separations larger than $\approx$ 5000 au. While it is outside of the scope of this work to identify the physical processes in which weakly bound WD+MS and WD+WD systems are disrupted, it is hoped that our 40\,pc sample can provide further insight.

Given \textit{Gaia} magnitude limit and lack of IR capabilities, our search is incomplete for substellar companions. We note the presence of at least one such confirmed system within the northern 40\,pc sample, WD 1422+095, which consists of a DA white dwarf at 33.4\,pc with a wide L4 brown dwarf companion separated by 120 AU \citep{becklin88}. The brown dwarf is not detected in \textit{Gaia} DR2.

Finally, at least three of our wide binaries are part of hierarchical triples. J043644.90+270951.52 is the outer companion of a spectroscopic binary of spectral type K2 \citep{holberg2013}. J163421.55+571008.87 is in a wide orbit around an eclipsing MS-MS binary of spectral type M \citep{toonen2017}. J170530.44+480312.36 is itself a double degenerate (see Section~\ref{sec:DD}) with J170530.97+480310.27 as a wide companion.
 
\subsection{Unresolved WD+MS}\label{binaryWD+MS}

From a search of the literature, we have found six unresolved WD+MS binaries that were missing from the selection of \citet{gentile2018gaia} and listed in Table~\ref{tab:A4}. \citet{hollands2018gaia} had identified none within the 20\,pc sample but a white dwarf companion to the M4.5V star LHS 1817 was recently discovered \citep{winters2020}. The five other unresolved systems are beyond 20\,pc. Based on binary population synthesis models that include observational selection effects, \citet{toonen2017} predict 3--8 unresolved WD+MS binaries within our surveyed volume. The six objects found are well within that range but our sample is very likely incomplete.

The majority of confirmed unresolved WD+MS systems at all distances have a low-mass M dwarf companion given the relative ease of identifying binarity from optical colours and spectroscopy \citep{rebassa16}. However, recent spectroscopic follow-ups such as in Paper~I have only focused on \textit{Gaia} sources consistent with single and double white dwarf parameters, and therefore the 40\,pc sample has incomplete spectroscopic coverage of unresolved WD+MS candidates. This is largely because of the high level of contamination of the local sample by \textit{Gaia} DR2 sources with incorrect or spurious astrometry \citep{gentile2018gaia}, which has proportionally much greater impact on less crowded regions of the HRD where composite spectra corresponding to unresolved binary systems are found. 

White dwarfs with more massive unresolved FGK type companions, as well as cool white dwarfs around M dwarfs, have remained elusive due to the main-sequence component usually dominating the optical flux \citep{holberg2013}. \citet{parsons2016} and \citet{rebassa17} identify more than a thousand WD+FGK candidates at all distances from their UV excess flux that they interpret as the likely presence of a white dwarf companion. However, the presence of a white dwarf has not been confirmed in any of their candidates within 40\,pc. Follow up observations with the Hubble Space Telescope (HST) has confirmed DA white dwarfs within 8 out of 9 of their systems outside 40\,pc, however.  Furthermore, this technique is only sensitive to white dwarfs warm enough to have detectable UV flux. Nevertheless, these searches could be updated in light of the newer \textit{Gaia} DR2 catalogue coupled with GALEX.

We note that our full sample of Table~\ref{tab:A1} contains only one spectral type (DB+dM) indicating an unresolved WD+MS system, J133601.82+482846.25, located at 36.9\,pc. However, the system is resolved with \textit{Gaia} DR2, and therefore classified as a wide binary system (Section~\ref{sec:wide-binary}).

Finally, in order to detect cool unresolved stellar companions, we have cross-matched our sample with 2MASS and WISE photometry, resulting in 481 and 429 matches, respectively \citep{gentile2020}. Given typical error bars of 0.05--0.10 mag on near-IR and IR magnitudes, we have only considered strong outliers ($>$0.4 mag) in $JHK$- and $W1$-band absolute magnitudes compared to the median value at a given $G_{\rm BP}-G_{\rm RP}$ colour. This resulted in seven candidates with a strong near-IR and IR excess. All of them are resolved wide systems in \textit{Gaia} and discussed in Section\,\ref{sec:wide-binary} but are only partially resolved or unresolved in 2MASS and WISE. Therefore, this search has not resulted in any additional unresolved WD+MS systems.

\subsection{Unresolved double white dwarfs}\label{sec:DD}

Twenty-six suspected or known double degenerates and low-mass white dwarfs are identified in Table~\ref{tab:DoubleWD} (see also comments in Table~\ref{tab:A1}). We discuss them in turn in this section.

\input{tables/Unres_double_WD.tex}

We initially flagged all objects for which {\it Gaia} $\log g < 7.72$ or the difference between the published spectroscopic and photometric $\log g$ values is greater than 0.5 dex. We have added J094846.64+242125.88 which does not fit the selection but is mentioned as a double DA+DAH in the literature \citep{liebert1993}. The published spectroscopic $\log g$ values in Table~\ref{tab:DoubleWD} as well as our photometric {\it Gaia} estimates are under the assumption of a single white dwarf and for illustrative purpose only. A low photometric $\log g$ under this assumption either suggests a low-mass white dwarf ($M \lesssim 0.45$ M$_{\odot}$) formed through binary evolution and with an unseen companion, or that two white dwarfs with normal $\log g$ values instead contribute to the total flux. We have reviewed all individual cases and removed those for which a spurious spectroscopic mass, e.g. because of magnetic fields or low S/N, is the most likely explanation. Very cool white dwarfs ($T_{\rm eff} < 4500$\,K) have a low mass problem (see Fig.~\ref{fig:logg_v_teff}), hence we have removed all candidates below this temperature. Our list of double degenerate candidates is therefore incomplete, both because of the initial selection of \citet{gentile2018gaia} which may have missed some unusually high luminosity (low-mass) sources and our secondary selection based on white dwarf parameters. However, the large majority of the objects in Table~\ref{tab:DoubleWD} are strong double dwarf candidates with no obvious alternative explanation, or already confirmed. Their mean photometric $\log g$ is 7.46, well below the average of $\log g \approx 8.0$ for the entire sample.

Only two candidates are within 20\,pc and were previously discussed in \citet{hollands2018gaia}. The increase in the number of double degenerate candidates is consistent with the increase in volume, given the low number statistics of the 20\,pc sample. Furthermore, only two of the candidates are from newly identified {\it Gaia} white dwarfs (J023117.04+285939.88 and J192359.24+214103.62). 

Seven objects are already confirmed multiple systems. There is a quadruple system (J010349.92+050430.57) consisting of a close double degenerate (1.2 or 6.4 hr period) with HD 6101, a wide dK3+dK8 resolved pair with 0.5$^{\prime\prime}$ separation \citep{maxted2000a,toonen2017}. The main sequence pair appears to have a disrupted and incorrect \textit{Gaia} DR2 parallax measurement, hence the system is not part of our sample of wide binaries including a white dwarf (Section~\ref{sec:wide-binary}). Our only triple WD system is the previously known J170530.44+480312.36, which consists of a double degenerate with a 0.15 hr period \citep{maxted2000b} and a wide white dwarf companion (see Section~\ref{sec:wide-binary}). Furthermore, five of the double degenerates  have been confirmed in the literature, either from radial velocity variations, double cores in Balmer lines at high-resolution or a composite spectrum. This includes the low-mass white dwarf J102459.83+044610.50 with radial variations over a 1.16 hr orbital period \citep{brown2011}.

At least eight of our double degenerate candidates have been discussed as such in the literature, including the two objects within 20\,pc. However, 10 other objects have no explicit identification as double degenerate candidates in the literature. Two of these are newly discovered white dwarfs from {\it Gaia} while eight objects had no reliable prior parallax measurements that would have been necessary to flag them as double degenerate candidates.

Of the only two confirmed low-mass white dwarfs, J094639.07+435452.24 does not have a confirmed companion although it likely formed through binary evolution \citep{brown2011}. Both low mass white dwarfs have $\log g$ = 7.50--7.70, based on either the photometric or spectroscopic technique, corresponding to $M$ = 0.35-0.45 \Msun. We do not have evidence of any extremely low-mass white dwarf \citep{kawka2020b} in our sample. However, some of the double degenerate candidates described above could also harbour a low-mass white dwarf ($\log g$ = 6.88--7.72 according to photometric values in Table~\ref{tab:DoubleWD}), especially in those cases where no spectroscopic $\log g$ value is available to confirm that the photometry is over-luminous compared to model predictions.

The binary population synthesis models of \citet{toonen2017} predict 5--33 unresolved WD+WD binaries within our surveyed volume, well within the range observed. We hope that our improved number statistics will help to further constrain binary population synthesis models, for which one of main source of uncertainty comes from the physics of the common-envelope phase.

\section{DISCUSSION}\label{sec:discussion}

\subsection{Space density}

The local space density of white dwarfs has historically been estimated using the 13 or 20\,pc volume-complete censuses \citep{sion2009white,giammichele2012know,holberg201625,hollands2018gaia} or larger samples corrected for completeness and Galactic structure effects \citep[see, e.g.,][]{munn2017,jimenez2018}. Based on the \gaia\ DR2 20\,pc sample and a careful consideration of its distance-dependent completeness, \citet{hollands2018gaia} derive a space density of (4.49 $\pm$ 0.38) $\times$ 10$^{-3}$\,pc$^{-3}$. That estimate is considerably larger than the value of (4.15 $\pm$ 0.35) $\times$ 10$^{-3}$\,pc$^{-3}$ found just by accounting for the 139 white dwarfs or double degenerates detected in \textit{Gaia} DR2 in the same volume. They established that \textit{Gaia} DR2 misses known objects at short distances, but is close to complete for white dwarfs near 20\,pc.

\citet{gentile2018gaia} found that Galactic structure effects, namely the density scale height, become increasingly significant for distances beyond $\approx$ 20\,pc. They determined that the 100\,pc white dwarf sample is best fit with an age-averaged density scale height of $\approx$ 250\,pc, assuming that \textit{Gaia} DR2 completeness does not change significantly within that distance \citep[see also][]{torres19}. While the 40\,pc sample is closer to the Galactic disc, selecting only the northern hemisphere amplifies Galactic effects. The northern sample favours Galactic latitudes that are further away from the Galactic plane than the southern hemisphere. Assuming a density scale height of 250 pc, this results in a space density 2.6 per cent lower in the northern 40\,pc sample compared to the full 20\,pc sample.

The actual numbers of systems including a white dwarf detected in \textit{Gaia} DR2 within 20\,pc (139) and within the northern 40\,pc sample (524) lead to space densities in agreement within 1$\sigma$. This is largely a consequence of the relatively large error on number statistics for the much smaller 20\,pc sample. However, this does not account for the fact that the {\it Gaia} detection rate is not constant as a function of distance. Given that the pre-\textit{Gaia} northern 40\,pc sample was at most $\approx$ 80 per cent complete \citep{limoges2015}, it is difficult to establish a robust list of white dwarfs missing from {\it Gaia} DR2. In other words, it is not sufficient to correct only for the known missing white dwarfs from Table~\ref{tab:A4} as we also need to include objects that are missing both from earlier samples and {\it Gaia} DR2. For this reason we consider that it is premature to update the white dwarf space density with larger volume samples. More work is also needed to understand the Galactic distribution and scale height of white dwarfs as a function of mass and age.

\subsection{Mass distribution}

\begin{figure}
	\includegraphics[width=\columnwidth]{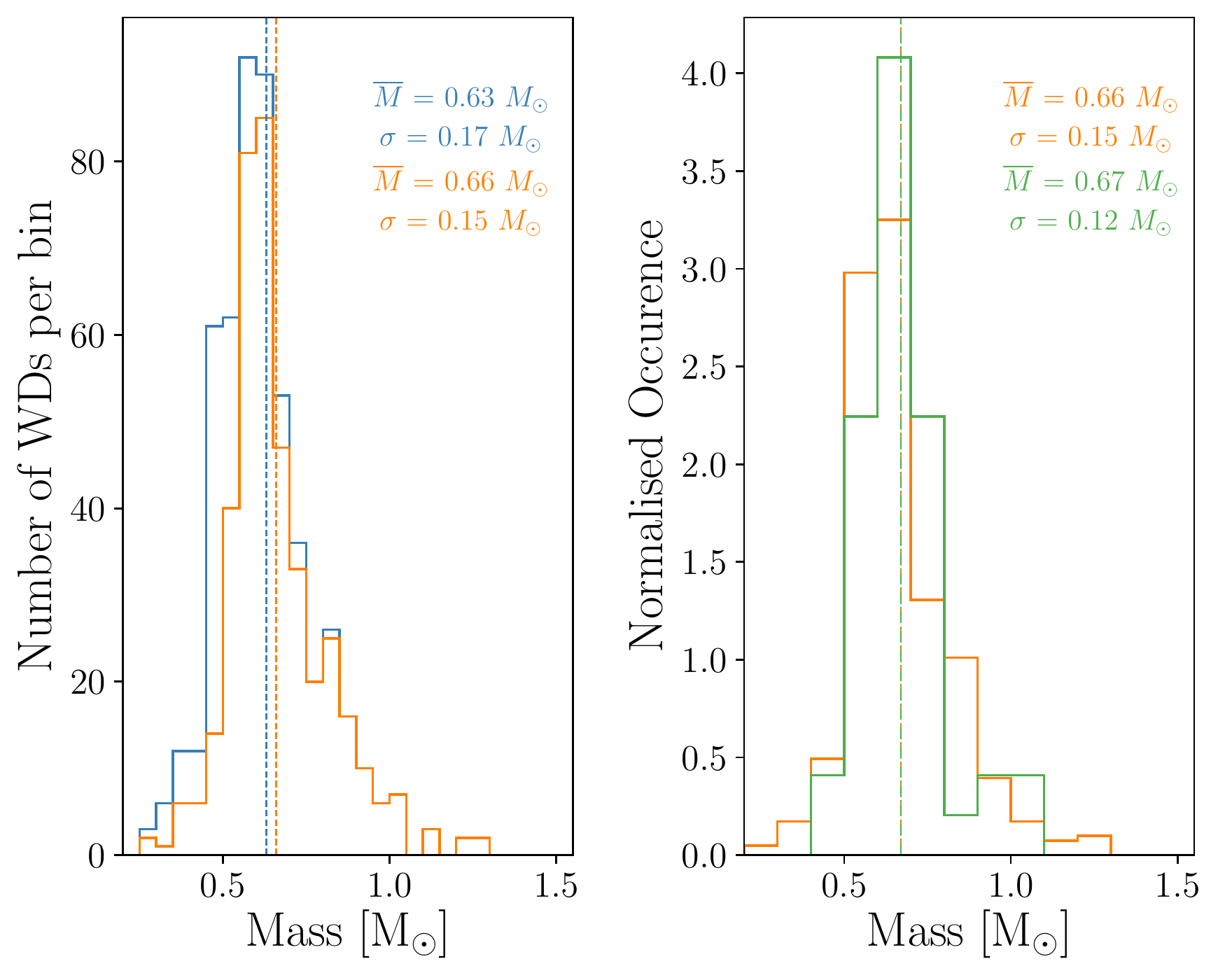}
    \caption{\textit{Left panel:} Mass distribution of the northern 40\,pc sample (blue) with the reduced sample of objects with \teff\,$>$\,$5000$\,K shown in orange ($\overline{M}_\textrm{40\,pc} = 0.66$\,\Msun\ and $\sigma_\textrm{40\,pc}$ = 0.15\,\Msun). \textit{Right panel:} Similar to left, a normalised distribution with the northern 40\,pc sample (\teff\,$>$\,$5000$\,K) in orange, and the full 20\,pc (\teff\,$>$\,$5000$\,K) sample in green ($\overline{M}_\textrm{20\,pc}$\,$=$\,0.67\,\Msun\  and $\sigma_\textrm{20pc}$ = 0.12\,\Msun.)}
    \label{fig:massdist}
\end{figure}

\begin{figure}
	\includegraphics[width=\columnwidth]{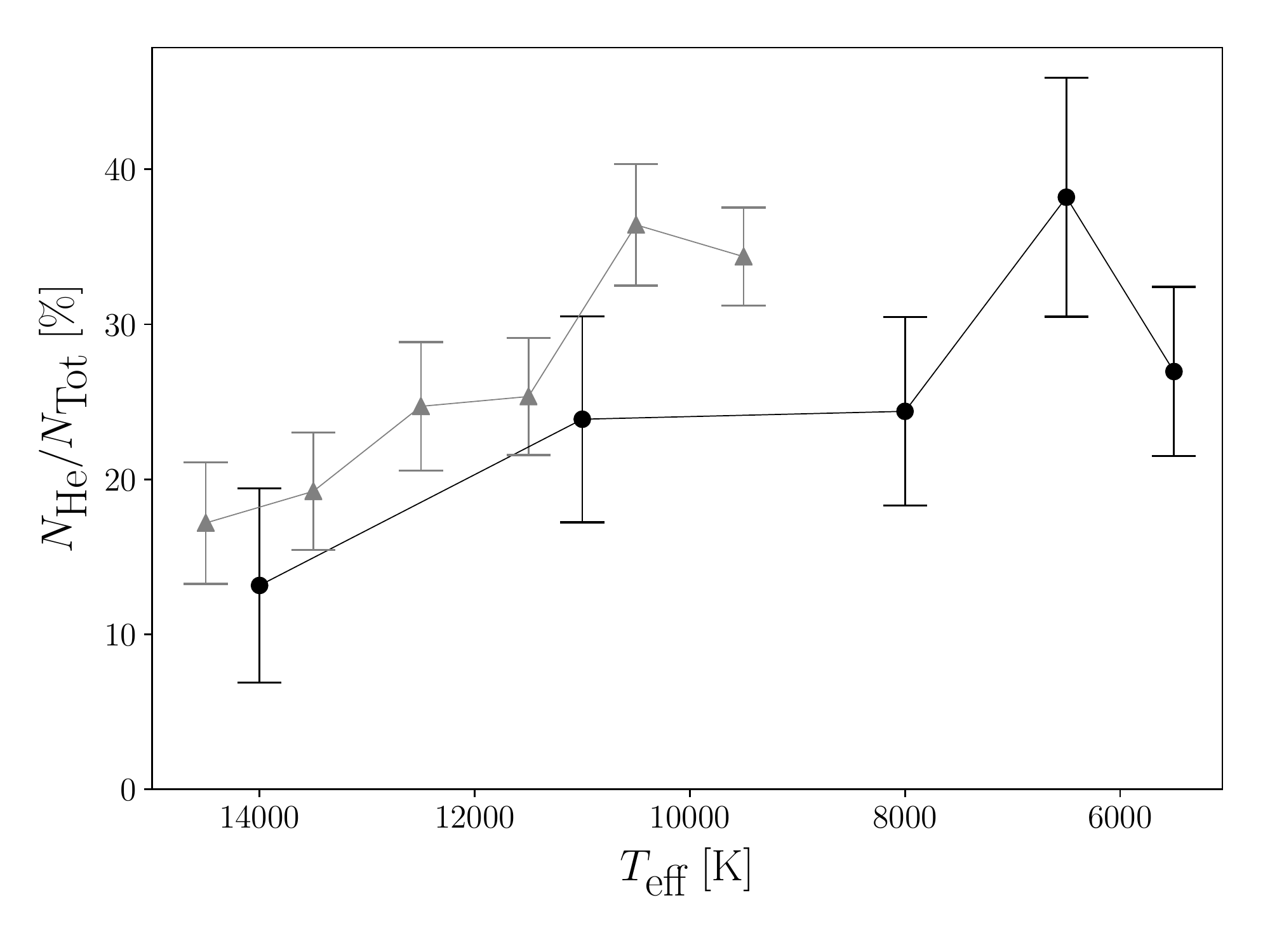}
    \caption{Percentage of He-rich atmosphere white dwarfs (DB, DC, DZ, DQ and He-rich DA) to the total as a function of effective temperature, \teff. Data from this work is shown in black, whilst the He incidence from \citet{cunningham2020} is shown in grey.}
    \label{fig:herichratio}
\end{figure}

In Fig. \ref{fig:massdist} we show the photometric mass distribution for the full 40\,pc northern sample in blue and for \teff\ > 5000 K in orange. Similarly to \citet{hollands2018gaia} we present the reduced distribution due to the low $T_\textrm{eff}$ regime being affected by the systematic decrease in $\log g$ (see Section\,\ref{sec:sample}). The mass distribution of \citet{hollands2018gaia} is also shown alongside ours, with the full sample shown in green, and the reduced sample in orange. Alongside the canonical peak located at 0.6\,\Msun, our sample does not show any significant secondary feature in the mass distribution \citep{tremblay2016,hollands2018gaia,temmink2019}. Both the 20\,pc and 40\,pc samples have a similar mean mass value of 0.66 -- 0.67\,\Msun\ (\teff\,$>$\,$5000$\,K). The astrophysical implications of the absolute mean mass value directly depend on the accuracy of {\it Gaia} photometric calibration \citep{tremblay2019,genest-beaulieu2019}. Nevertheless, using Pan-STARRS we find a mean mass of 0.65\,\Msun\ for the 40\,pc sample, in good agreement with the \textit{Gaia} result and in the same range as previous estimates for the local volume sample \citep{tremblay2016}.

We find the same mean mass of 0.66\,\Msun\ for both H- and He-rich white dwarfs (\teff\,$>$\,$5000$\,K). This may be a coincidence from using He-atmosphere models with a fixed H/He = 10$^{-5}$ trace abundance, which is not expected to fully represent the effects of hydrogen and metals in these atmospheres. Given our temperature cut, the majority of our He-rich atmospheres fall within the range where a bifurcation appears in the HRD when using pure-He models \citep{gentile2018gaia}. Nevertheless, \citet{tremblay2019} have shown that warmer DB and DBA white dwarfs in larger magnitude limited samples have a similar mean mass to DA white dwarfs. Since DB, DBA and DA white dwarfs are thought to be the progenitors of the cool He-rich atmospheres in our sample \citep{cunningham2020}, it is expected that they should also have the same mean mass as DA white dwarfs. This suggests that our adopted fixed trace hydrogen abundance leads to a sound astrophysical result, although the different mean masses for DC, DQ and DZ white dwarfs as described below is still a source of concern.

The 17 DQ white dwarfs have a mean mass of 0.62~\Msun, which is lower than the average. Since cool DQ white dwarfs have broad optical absorption bands, the difference in mass could be explained by the fact that we use mixed H/He atmospheres without carbon for these objects. For 20 DZ white dwarfs we find a slightly lower mean mass ($\overline{M}$ = 0.63\,\Msun, $\overline{M}_{\rm >5000\,K}$ = 0.64\,\Msun) than other He-atmospheres, which could be the result of our neglect of metals in model atmospheres. This effect is even more pronounced for 22 DAZ white dwarfs ($\overline{M}$ = 0.58\,\Msun, $\overline{M}_{\rm >5000\,K}$ = 0.59\,\Msun), in which the presence of metals is not expected to contribute to any significant opacity that would lead to a systematic effect on mass determinations. The mean mass difference could be caused by selection biases in detecting metals in DA white dwarfs. Lower mass white dwarfs are more luminous for a volume-limited sample and have brighter apparent magnitudes on average, possibly facilitating metal detection. Nevertheless, the mean mass difference of 0.07\,\Msun\ between DA and DAZ is large and unlikely to be fully explained by this bias. A Kolmogorov-Smirnov test gives a p-value of 1.9 per cent, corresponding to the likelihood that random fluctuations can explain the mean mass difference. This confirms that the mean mass difference is statistically significant but we cannot fully reject a random fluctuation. This could suggest either that planet formation occurs more frequently in lower mass stars or that planetary debris scattering onto central white dwarf occurs less frequently in remnants with higher mass progenitors \citep{veras2020}.

\subsection{Spectral evolution}

Figure~\ref{fig:herichratio} shows the percentage of He-rich objects as a function of $T_{\rm eff}$. The small number of warm white dwarfs in the 40\,pc sample does not allow to have more than 1 bin above 12\,000\,K. Even below that temperature, the error bars due to number statistics remain larger than the small observed fluctuations, suggesting that larger volume samples will be needed to fully address spectral evolution \citep{macdonald1991,bergeron2001,tremblay2008,blouin2019,cunningham2020}.

Below $T_{\rm eff} = 20\,000$\,K, the only processes that have been invoked to change white dwarf surface abundances are convective mixing of the underlying helium layer with the top hydrogen layer \citep{rolland2018,cunningham2020}, accretion of planetary debris \citep{ngf2017} or convective dredge-up of carbon \citep{koester82}. Only the first scenario of convective mixing is thought to result in a change of the dominant atomic species in the atmosphere, namely a transition from H- to He-dominated (DA to DB(A) or DC). Hence, convective mixing is currently the only reasonable scenario to explain variations in Fig.~\ref{fig:herichratio}.

Our results can be compared in Fig.~\ref{fig:herichratio} to the study of \citet{cunningham2020} who employed SDSS, GALEX and \textit{Gaia} photometry to study spectral evolution in the range 9000\,K $\leq T_{\rm eff} \leq$  20\,000\,K from the strength of the Balmer jump in the 133\,pc volume-limited sample. They found a He-rich percentage with respect to total number of white dwarfs of 18 $\pm$ 3 per cent
at 13\,000--15\,000\,K and 34 $\pm$ 3 per cent for the bin in the range 9000--10\,000\,K. This increase by a factor of $\approx$ 2 is consistent with the picture provided by the 40\,pc sample. The He-rich fraction in the 40\,pc sample is consistently lower compared to the results found in \citet{cunningham2020}, but still agrees to within 1-2$\sigma$ for any temperature bin given the small number statistics.

The spectral evolution for $T_{\rm eff} < 8000$\,K was also studied by \citet{blouin2019} using a spectral type identification method similar to the one employed in this work, albeit with a different sample incomplete in volume but likely representative of the local white dwarf population. They also used different model atmospheres. Neglecting the range $T_{\rm eff} < 5000$\,K for which we did not adopt atmospheric compositions, they find a He-rich percentage of $\approx$ 20--25 per cent in the range 5000-8000\,K, which is marginally lower than our average value of 30 $\pm$ 4 per cent. 

Both studies suggest an increase and subsequent decrease in the He-rich fraction around 6500\,K, although in our case the significance is only at the 1--2$\sigma$ level. The increase can be explained by the occurrence of convective mixing in white dwarfs with relatively thick hydrogen layers ($\log (M_{\rm{H}}/M_{\rm WD}) \sim -7$). But given that this process is non-reversible, there exists no obvious physical explanation for the decrease in He-rich fraction below 6500 K. Similar behaviour has been observed in previous photometric and spectroscopic studies, with an apparent deficit of He-rich objects between $\approx$ 6000--5000\,K being coined the ``non-DA gap" \citep{bergeron1997,leggett98,bergeron2001}. This observation was investigated by \citet{chen12} who proposed that the non-DA gap could be explained by convective mixing in white dwarfs where the convection zone is coupled to the degenerate core. Convective coupling occurs when the base of the convection zone grows deep enough to reach the degenerate core. From evolutionary models this occurs in white dwarfs with effective temperatures of $\approx6000 \pm 300$\,K (see Fig.~6 of \citealt{fontaine13}) and results in surface layers which are strongly coupled to the central thermal reservoir via an almost adiabatic convection zone. In such a scenario, \citet{chen12} hypothesised that following convective mixing the surface layers should experience an increase in effective temperature of $\approx 500$\,K. Employing a Monte Carlo approach they found that the non-DA gap arose as a natural consequence of convective mixing in white dwarfs with effective temperatures $\approx6000$\,K. Whilst this model has not been well-constrained by observations, it provides an explanation for the feature we observe at 6500\,K (Fig~\ref{fig:herichratio}). In the future larger volume-limited samples will be needed to ascertain the statistical significance of this bump.

\subsection{Magnetism}\label{sec:mag}

\begin{figure}
	\includegraphics[width=\columnwidth]{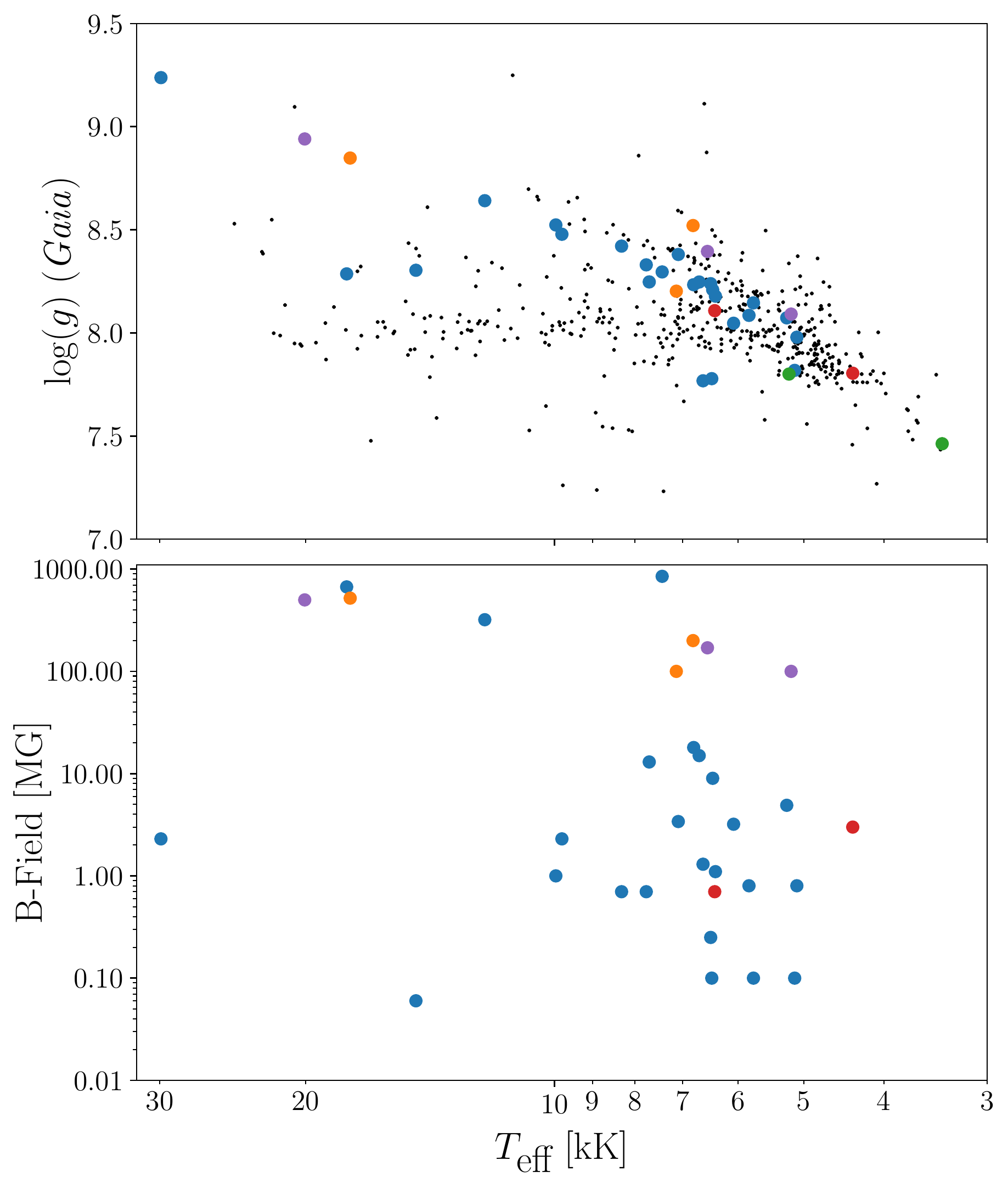}
    \caption{\textit{Top Panel:} Atmospheric parameters of magnetic white dwarfs of different spectral types compared to the full 40\,pc sample (black points). \textit{Bottom Panel:} Magnetic field strength as a function of $T_{\rm eff}$ for different spectral types. In both panels, DAH/DAP/DAZH are shown in blue, DCP/DBP in orange, DQP in green, DZH in red, and DX/DXP in purple.}
    \label{magnetic11}
\end{figure}

Table~\ref{tab:magnetic} summarises all white dwarfs with a spectral type indicating magnetism. It does not include WD\,2150+591, a DAH white dwarf at $8.34$\,pc which does not have an astrometric solution in \textit{Gaia} DR2 (see Table~\ref{tab:A4}). The top panel of Fig.~\ref{magnetic11} shows the distribution of magnetic white dwarfs in the $T_{\rm eff}$ versus $\log g$ diagram. It excludes the DQpecP J101141.58+284559.07 for which we have no reliable \textit{Gaia} atmospheric parameters. The results suggest that magnetic white dwarfs are more massive than the average. We also observe that many magnetic white dwarfs follow the crystallisation sequence (see Section \ref{sec:crystal}).

Our volume-limited sample allows to look at statistics of field strength versus cooling age as done by \citet{landstreet2019} for the 20\,pc sample. The results are shown in the bottom panel of Fig.~\ref{magnetic11}  for the northern hemisphere 40\,pc sample. This excludes DQP J123752.23+415624.69 for which the magnetic field strength is unclear. We observe no obvious correlation between field strength and temperature, as was found by \citet{landstreet2019}.

\input{tables/magnetic_objs.tex}

\begin{figure}
	\includegraphics[width=\columnwidth]{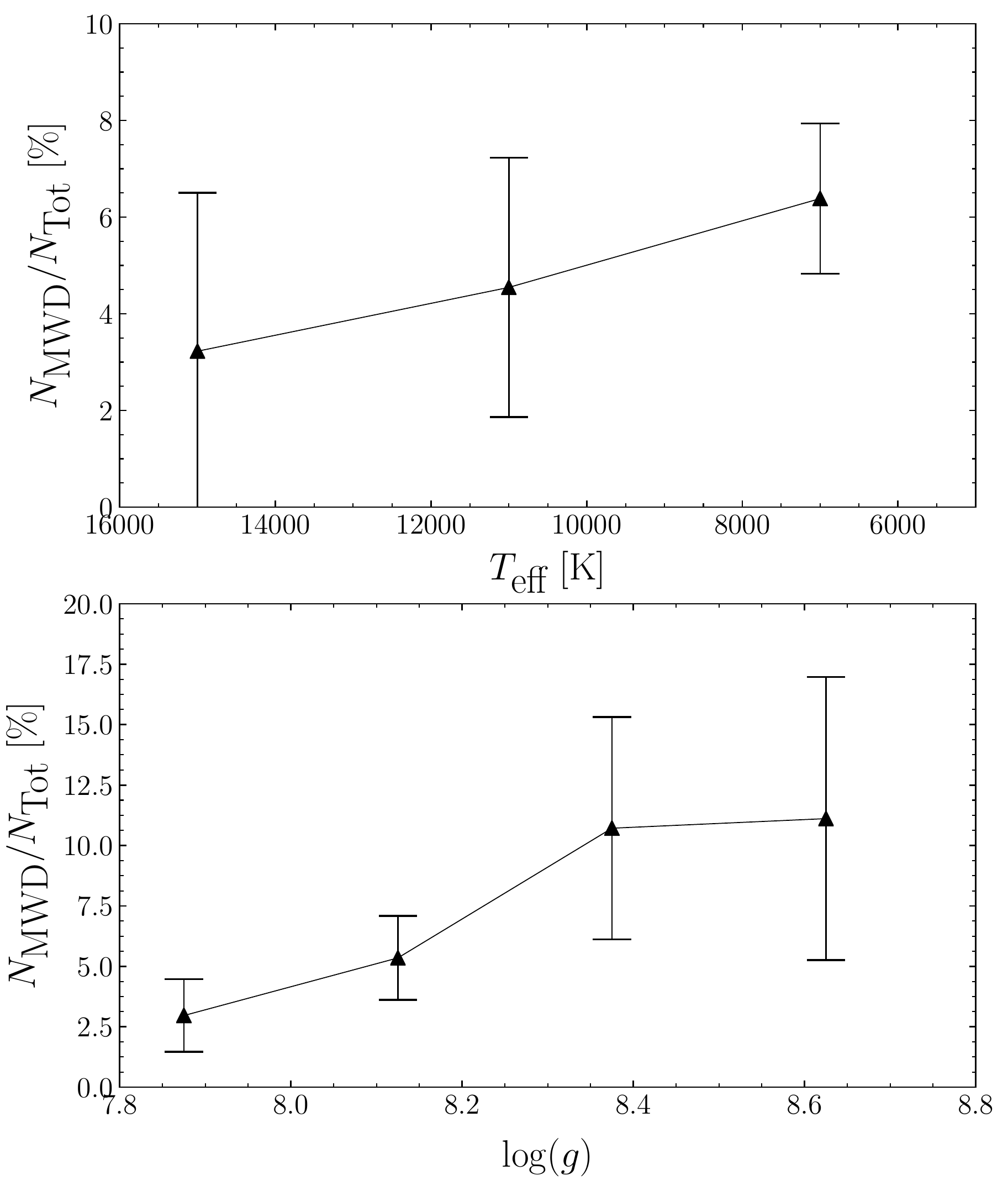}
    \caption{\textit{Top Panel:} Magnetic incidence as a function of $T_{\rm eff}$ for H-atmosphere white dwarfs. \textit{Bottom Panel:} Magnetic incidence as a function of $\log g$ for H-atmosphere white dwarfs.}
    \label{magnetic33}
\end{figure}

For statistics as a function of mass and temperatures, we limit ourselves to DA white dwarfs with $T_{\rm eff} > 5000$\,K for which Zeeman splitting can be detected. This is because most helium-rich atmospheres are of DC spectral type, for which magnetic field detection is difficult. Even spectropolarimetric observations are orders of magnitude less sensitive to magnetic fields in DC white dwarfs compared to DA spectral type  \citep{landstreet2019}. Fig.~\ref{magnetic33} shows magnetic fraction histograms with $T_{\rm eff}$ and $\log g$ as independent variables for hydrogen-atmosphere white dwarfs. For both parameters, number statistics are relatively poor given that the curves are based on only 24 magnetic hydrogen-atmosphere white dwarfs within 40\,pc. It is clear that larger samples will be needed to confirm any correlation of magnetism with temperature and mass, in addition to potential biases against the detection of small magnetic fields. We note that since more massive white dwarfs are intrinsically fainter, the tentative increase in incidence as a function of $\log g$ observed in the bottom panel of Fig.~\ref{magnetic33} can not easily be explained by observational biases. Overall, the 40\,pc sample provides strong evidence that magnetic white dwarfs are more massive than the average, with $\overline{M} = 0.75$ M$_{\odot}$ for 24 DAH, DAP and DAZH versus $\overline{M} = 0.66$ M$_{\odot}$ for 278 non-magnetic DA and DAZ white dwarfs above 5000\,K. A Kolmogorov--Smirnov test gives a p-value of much less than one per cent, meaning we can reject this being a random fluctuation.

\begin{figure}
	\includegraphics[width=\columnwidth]{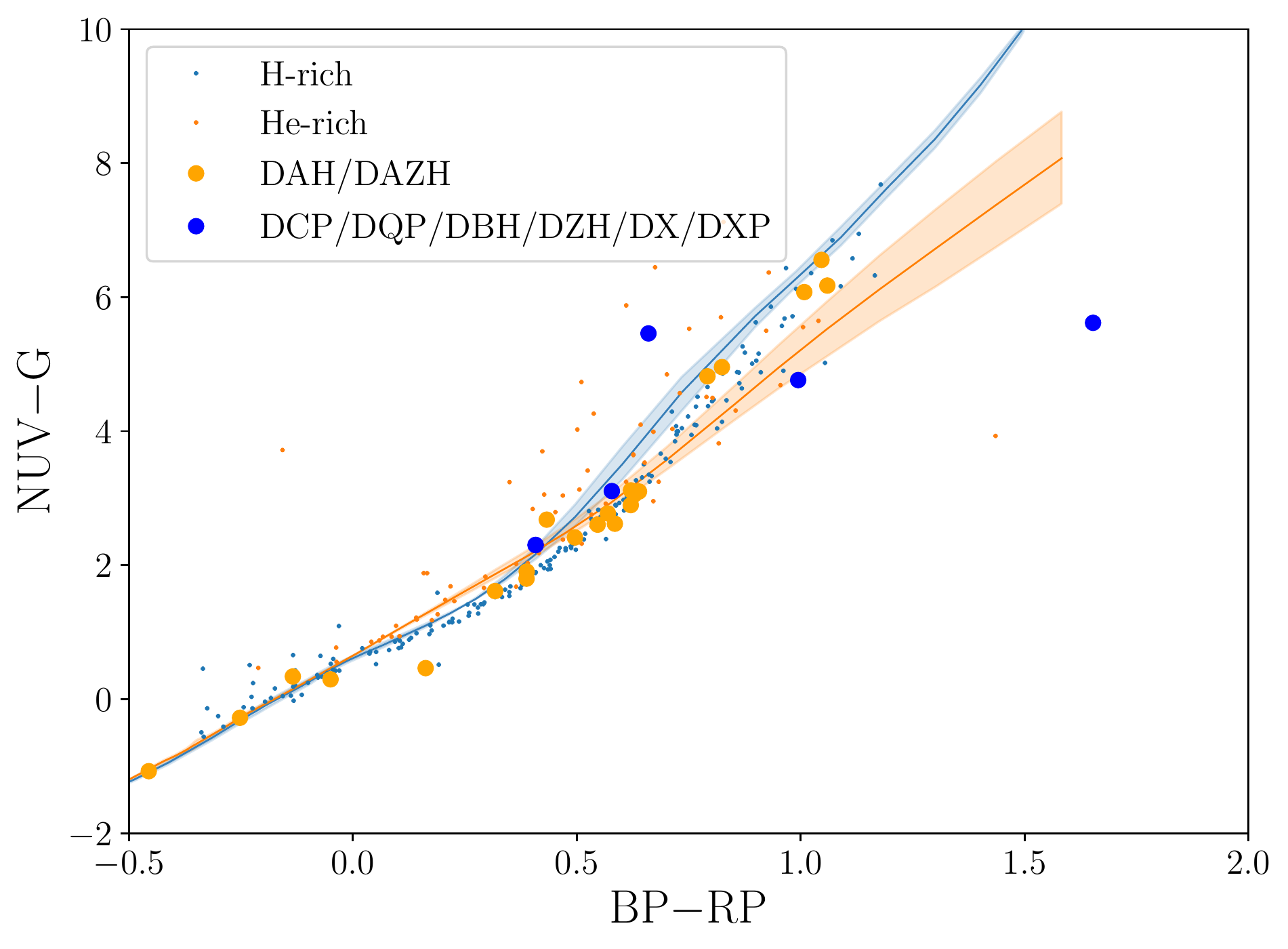}
    \caption{Colour-colour optical \textit{Gaia} vs. GALEX NUV diagram highlighting magnetic white dwarfs with H- (blue) and He-rich (orange) composition. The shaded region indicates pure-H and pure-He model predictions for  $7.5<\log g<8.5$ with $\log g=8.0$ shown with a darker line.}
    \label{magnetic5}
\end{figure}

\begin{figure}
	\includegraphics[width=\columnwidth]{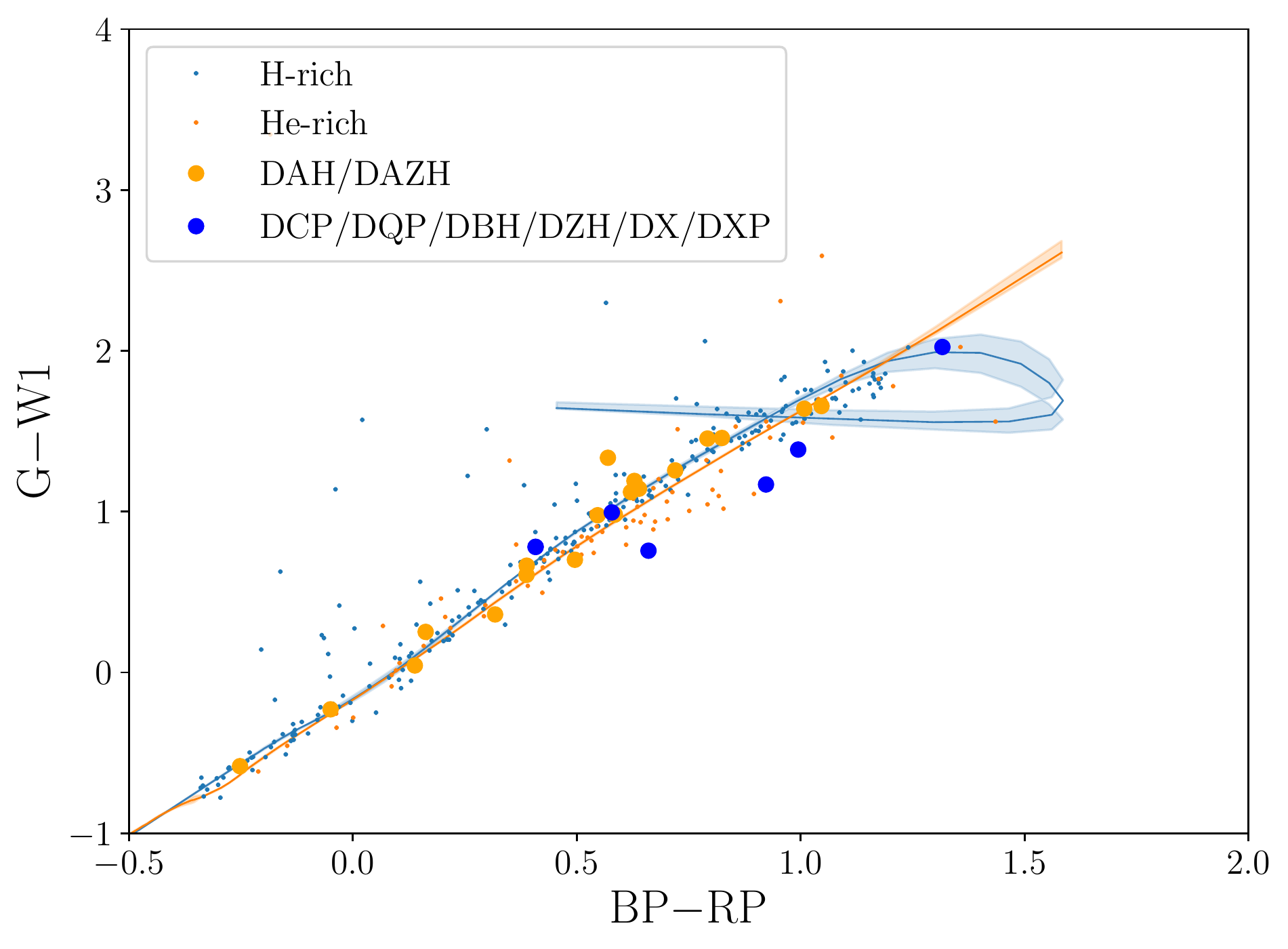}
    \caption{Colour-colour optical \textit{Gaia} vs. WISE W1 near-IR diagram highlighting magnetic white dwarfs with H- (blue) and He-rich (orange) composition. The shaded region indicates pure-H and pure-He model predictions for  $7.5<\log g<8.5$ with $\log g=8.0$ shown with a darker line.}
    \label{magnetic6}
\end{figure}

We have employed non-magnetic models to fit \textit{Gaia} colours of magnetic white dwarfs. This could lead to a systematic effect on their atmospheric parameters. However, we note that they align with the same crystallisation sequence as non-magnetic white dwarfs. This is expected as magnetic fields of $B<1000$ MG have little influence on the cooling process \citep{tremblay2015}, suggesting that  \textit{Gaia} derived atmospheric parameters of magnetic white dwarfs are accurate to at least a few percent. To obtain a better estimate of the accuracy of the atmospheric parameters for magnetic white dwarfs, Figs.~\ref{magnetic5}-\ref{magnetic6} show GALEX-\textit{Gaia} and \textit{Gaia}-WISE colour-colour diagrams, respectively. It demonstrates that for the vast majority of the magnetic white dwarfs in the sample, the shape of the energy distribution is empirically similar to non-magnetic white dwarfs. In other words, magnetic white dwarfs can be fitted with a single white dwarf model from the near-UV to the near-IR. Since this range covers most of the emergent flux, it suggests that \textit{Gaia} $T_{\rm eff}$ and mass values for magnetic white dwarfs are similarly accurate as those of non-magnetic white dwarfs. We conclude that the previously claimed higher-than-average mass values of magnetic white dwarfs \citep{ferrario2015,kawka2020} is a robust result. 

\subsection{Crystallisation}\label{sec:crystal}

Table~\ref{tab:crystallisation} provides statistics for white dwarfs roughly defined to be on the crystallisation sequence \citep{tremblayNature} from

\begin{multline}
\log g_{\rm Gaia} > \frac{[(T_{\rm eff, Gaia}-5000)/1000]^{0.95}}{10}+8.00~{\rm and}\\
\log g_{\rm Gaia} < \frac{[(T_{\rm eff, Gaia}-5000)/1000]^{0.95}}{4}+8.00~{\rm and}\\
6000 < T_{\rm eff, Gaia} \rm{[K]} < 12\,000~.
\label{eq:cry}
\end{multline}

\begin{figure}
	\includegraphics[width=\columnwidth]{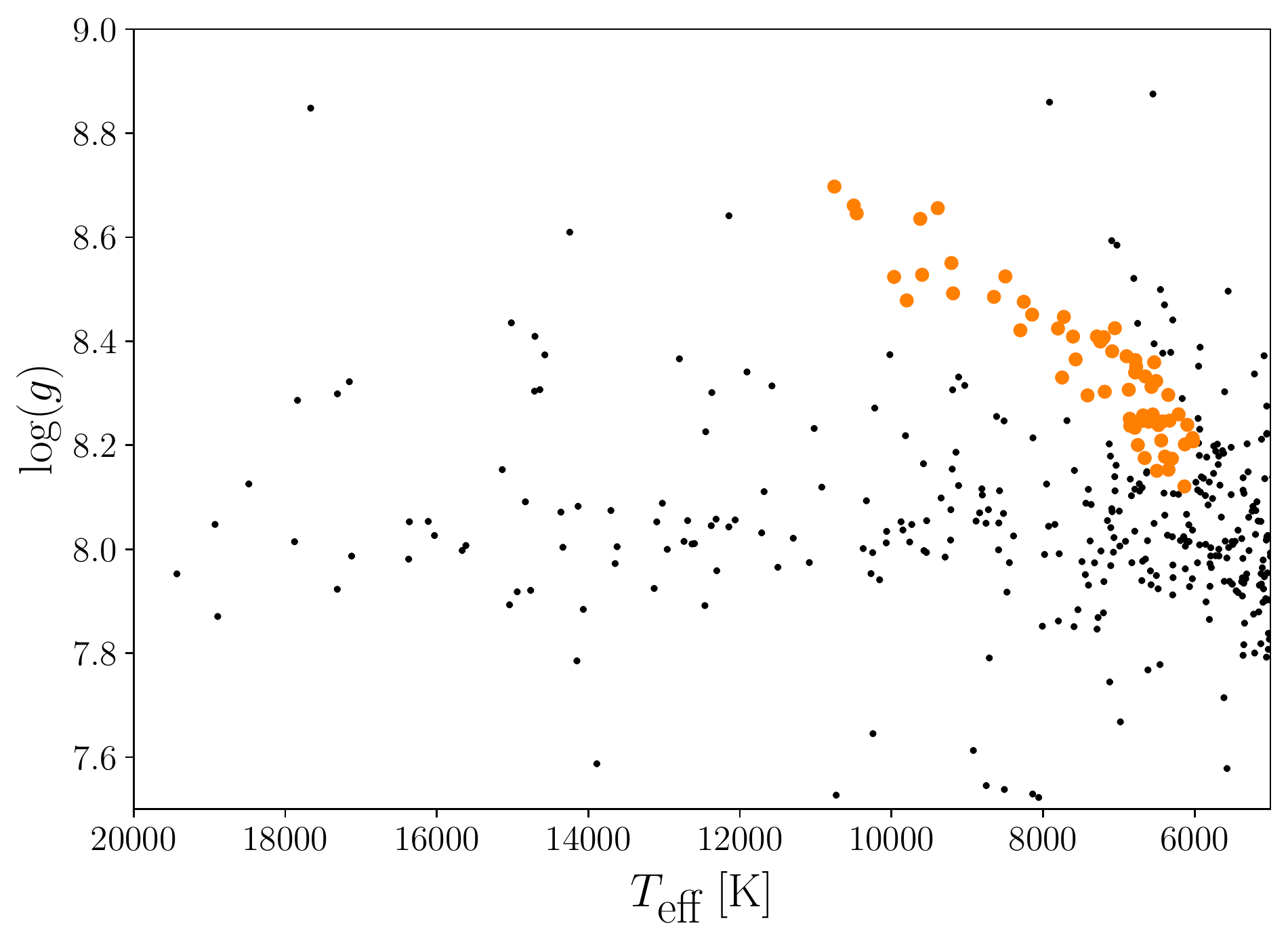}
    \caption{Empirically selected crystallisation sequence based on Eq.~(\ref{eq:cry}) and the \textit{Gaia} $\log g$-temperature distribution.}
    \label{fig:cryst}
\end{figure}

{\noindent}This empirically selected region of over-density is shown in Fig.~\ref{fig:cryst}. We apply a lower temperature limit because there is no distinct  crystallisation sequence below 6000\,K, as it merges with the peak in the $\log g$ distribution, e.g. the large majority of these cool white dwarfs could have started the crystallisation process. The upper temperature limit is applied because it is difficult to empirically isolate a crystallisation sequence above that temperature, i.e. only four hot and massive white dwarfs are potentially on the sequence. By applying the upper temperature cut, we also have average temperatures that are more similar for crystallising white dwarfs and those not on the sequence. We note that this experiment is different to the comparison of white dwarfs that have a liquid interior and those that have crystallised. Only a handful of massive white dwarfs have a temperature above 6000\,K and are likely to be fully crystallised, but in the present analysis those are included in the bin of objects not currently crystallising.

\begin{table}
\centering
\caption{Statistics of Crystallising and Non-Crystallising White Dwarfs (6000\,K $<$ \teff\ $<$ 12\,000\,K)}
\begin{tabular}{@{}lll@{}}
\toprule
 -- & Crystallisation Seq. & Not on Seq. \\
\hline
$\overline{\log g}$ [cm/s$^2$] & 8.37 & 8.13 \\
$\overline{T_{\rm eff}}$ [K] & 7460 & 7800 \\
$\overline{\nu_{\perp}}$ [km/s] & 40.3 & 39.8 \\
N(DQ)/$N_{\rm tot}$ & 1.6\% $\pm$ 2\% & 6.4\% $\pm$ 2\% \\
N(DA)/$N_{\rm tot}$ & 79\% $\pm$ 15\% & 70\% $\pm$ 7\% \\
N(Magnetic)/$N_{\rm tot}$ & 18\% $\pm$ 6\% & 8.5\% $\pm$ 2\% \\
\bottomrule
\end{tabular}
\label{tab:crystallisation}
\end{table}

Table~\ref{tab:crystallisation} demonstrates that white dwarfs on the crystallisation sequence are in most measurable quantities very similar to white dwarfs that are not yet at that stage or have already solidified. In particular, white dwarfs on the crystallisation sequence are overwhelmingly of DA spectral type as outlined in \citet{tremblayNature}. By construction white dwarfs with $T_{\rm eff} > 6000$\,K on the crystallisation are more massive than the average by about 0.12\,\Msun. Hence, any difference in their properties could be explained by their past evolution at these characteristic higher masses, and not necessarily by the crystallisation process itself. In particular, we detect a magnetic fraction that is marginally higher on the crystallisation sequence. Considering that the fraction of magnetic white dwarfs increases for the full sample as a function of mass (see Section\,\ref{sec:mag}), we speculate that there is no obvious causal link between magnetic field generation and crystallisation given the current sample size. 

We find no significant difference between kinematic properties of crystallising white dwarfs and those that are not. This suggests that below $M \lesssim 1.0$\,\Msun, there is no need to invoke a population of WD+WD mergers to explain the properties of crystallising white dwarfs \citep{cheng2019}. This is a much different picture to the regime $M > 1.08$\,\Msun\ studied in \citet{cheng2019} and for which there is a significant difference between the kinematics of white dwarfs on the crystallisation sequence. In the northern 40\,pc sample, only 6 objects have $M > 1.08$\,\Msun\ as a consequence of the steep initial mass function. Assuming that 20 per cent of those come from WD+WD mergers as suggested by \citet{cheng2019}, this would result, on average, in only a single example in our surveyed volume. Hence, such high-mass merger population would not produce any detectable signal in our 40\,pc sample. The fraction of white dwarfs that come from mergers at lower masses is still an open issue which is difficult to quantify with our sample as these objects are likely to have measurable properties that are similar to white dwarfs that have evolved through single star evolution \citep{temmink2019}.

\subsection{Ultra-cool white dwarfs}

\begin{figure}
	\includegraphics[width=\columnwidth]{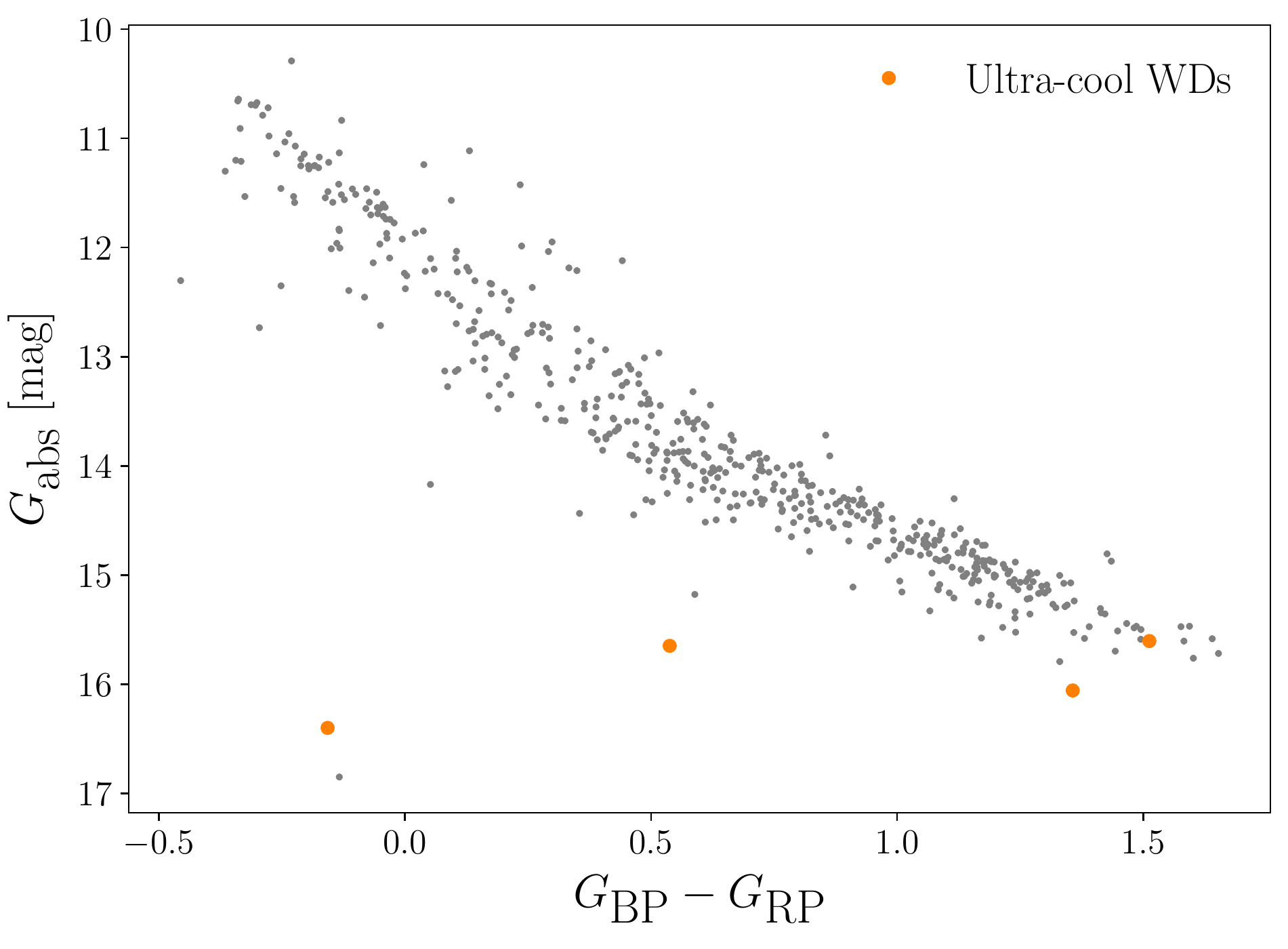}
    \caption{Position in the \textit{Gaia} HRD of the four known ultracool white dwarfs with strong collision induced absorption and within the northern 40\,pc sample (orange points). The full northern 40\,pc white dwarf sample is shown in gray.}
    \label{fig:UCWD}
\end{figure}

Within our sample there are only four so-called ultra-cool white dwarfs ($\approx1$ per cent), whose main feature is strongly non blackbody-like optical and near-IR colours due to CIA \citep{blouin2017}. The location of these four ultra-cool white dwarfs in the \textit{Gaia} HRD is shown in Fig. \ref{fig:UCWD}. These are among the faintest in the sample, lying below $G_{\rm abs}=15.5$. However they do span a very wide range in \textit{Gaia} colours, making it difficult to estimate the completeness of the current sample. The selection of \citet{gentile2018gaia} managed to recover all four of the brightest previously known ultra-cool white dwarfs, suggesting that few additional objects are missing within 40\,pc. Nevertheless, they lie in regions of the HRD where white dwarf identification is difficult due to a significant contamination from erroneous measurements. Two additional objects in the sample are possible ultra-cool white dwarfs, the cool DZ J192206.20+023313.29 (see Paper I) and J050600.41+590326.89 for which spectroscopy is missing (see Section~\ref{sec:missing}). Finally, we note that other objects in the sample show milder CIA in the near-IR. It is likely that this class of objects can not only be defined by surface temperature and chemical composition could also play a role \citep{kilic2020}.

Table \ref{tab:UCWD} shows the parameters of ultra-cool white dwarfs as found in the pre-\textit{Gaia} literature, although caution should be used due to the uncertain nature of earlier CIA opacity calculations \citep[see, e.g.,][]{mixedmodelpier2014}. Curiously only two of the four objects have halo-like kinematics, while the other two are consistent with the Galactic disc (see Section\,\ref{sec:kin}). This could suggest that ultra-cool white dwarfs do not form a homogeneous Galactic population.

\input{tables/UCWD.tex}

\section{Conclusions}\label{sec:conclusions}

We have presented a review of the \textit{Gaia} DR2 selected sample of white dwarfs within the 40\,pc northern hemisphere. This corresponds to the largest and most complete volume-sample of white dwarfs with low-resolution spectroscopic confirmation available so far, an increase of a factor of about four in number compared to the 20\,pc sample \citep{hollands2018gaia}. Our selection is based on the catalogue of \citet{gentile2018gaia} while we have gathered spectral types from the literature and a companion paper \citep{tremblay2020_WHT}. This results in a final sample of 524 white dwarfs, among which 521 have known spectral types. The existing observations are sufficient to determine the dominant atmospheric chemical constituent of individual white dwarfs (above $\approx$ 5000\,K), resulting in much better constraints on the atmospheric parameters compared to larger volume samples with low spectroscopic completeness. However, the varying quality of the spectroscopic signal-to-noise between observations prevents the derivation of robust absolute numbers of sub-types such as magnetic and metal-poluted white dwarfs. We note a systematic but moderate change in detection rates of sub-types with distance, which can only be explained as an observational bias.

We find that most properties of the northern 40\,pc sample, such as the mean mass, kinematics, and fraction of wide and unresolved binaries are similar to those found for the four-time smaller 20\,pc sample, but some new trends appear in more rare subtypes of white dwarfs. We find a significantly lower mean mass for metal-rich DAZ white dwarfs compared to their parent DA population, which can not easily be explained as a simple statistical fluctuation. We suggest that it may inform about planet formation as a function of stellar mass \citep{veras2020}. We find a significantly larger mean photometric mass for magnetic white dwarfs, which we determine is robust even though we rely on predicted magnitudes from non-magnetic models. Finally, the sample contains a notable sequence in $\log g$-$T_{\rm eff}$ space of white dwarfs currently undergoing crystallisation \citep{tremblayNature}, which appear otherwise typical in all other measurable quantities.

The main advantage of the 40\,pc sample is that of better number statistics, possibly enabling the study of science questions that were not possible to answer with smaller samples. One example is the study of the local stellar formation history by relying on the reasonably well constrained white dwarf ages \citep{tremblay2014,fantin2019}. The local white dwarf sample could also be combined with the full 40\,pc \textit{Gaia} stellar sample to study overall stellar evolution, especially as new spectroscopic observations become available.

\section*{Acknowledgements}
The research leading to these results has received funding from the European Research Council under the European Union's Horizon 2020 research and innovation programme n. 677706 (WD3D). This work presents results from the European Space Agency (ESA) space mission Gaia. Gaia data are being processed by the Gaia Data Processing and Analysis Consortium (DPAC). Funding for the DPAC is provided by national institutions, in particular the institutions participating in the Gaia MultiLateral Agreement (MLA). BTG has been supported by the STFC grant ST/T000406/1. ST acknowledge support from the Netherlands Research Council NWO (VENI 639.041.645 grants). ARM acknowledges  support  from  the MINECO under  the  Ram\'on  y  Cajal programme (RYC-2016-20254), the AYA2017-86274-P grant and the AGAUR grant SGR-661/2017. PI acknowledges financial support from the Spanish Ministry of Economy and Competitiveness (MINECO) under the 2015 Severo Ochoa Programme MINECO SEV--2015--0548. The work presented made extensive use of TOPCAT \citep{topcat}.

\section*{Data Availability Statement}
The data underlying this article are available in the Gaia and Pan-STARRS public archives.




\bibliographystyle{mnras}
\bibliography{wd40pc}


\appendix

\section{Online Tables}

\input{tables/TABLEA1.tex}
\input{tables/TABLEA2.tex}
\input{tables/TABLEA4.tex}
\input{tables/TABLEA3.tex}
\input{tables/TABLEA5.tex}


\bsp	
\label{lastpage}
\end{document}

%% file: tables/headers.tex
\begin{table*}
\centering
\caption{Format of the online catalogue which can be accessed at \href{http://deneb.astro.warwick.ac.uk/phrgwr/40pcTables/index.html}{this link}.}
\begin{tabular}{@{}llll@{}}
\toprule
Index & Column Name                   & Units & Description \\ \midrule
1     & WDJ Name                      & --    & WDJ + J2000 ra (hh mm ss.ss) + dec (dd mm ss.s), equinox and epoch 2000\\
2     & WD Name                       & --    & WD Name (for objects known before \gaia\ only)           \\
3     & Source ID                     & --    & Gaia DR2 source identifier            \\
4     & Parallax                      & mas   & Parallax of the source            \\
5     & Parallax\_Error               & mas   & Standard error of parallax            \\
6     & RA                            & deg   & Right ascension (J2015.5)           \\
7     & RA\_Error                     & mas   & Standard error of right ascension             \\
8     & DEC                           & deg   & Declination (J2015.5)            \\
9     & DEC\_Error                    & mas   & Standard error in declination           \\
10    & appG                          & mag   & Apparent $G$ magnitude            \\
11    & bp\_rp                        & mag   & $G_{\rm BP} - G_{\rm RP}$ colour index          \\
12    & SpT                           & --    & Spectral Type            \\
13    & Comp                          & --    & Composition (H for hydrogen dominated or He for helium dominated)     \\
14    & Gaia Teff                     & K     & Adopted \gaia\ effective temperature           \\
15    & Gaia Teff Error               & K     & Standard error of adopted \gaia\ effective temperature            \\
16    & Gaia log(g)                   & [cm/s$^2$]  & Adopted \gaia\ surface gravity          \\
17    & Gaia log(g) Error             & [cm/s$^2$]  & Standard error on adopted \gaia\ surface gravity          \\
18    & Pan-STARRS Teff               & K     & Pan-STARRS effective temperature            \\
19    & Pan-STARRS Teff Error         & K     & Standard error on Pan-STARRS effective temperature           \\
20    & Pan-STARRS log(g)             & [cm/s$^2$]  & Pan-STARRS surface gravity           \\
21    & Pan-STARRS log(g) Error       & [cm/s$^2$] & Standard error on Pan-STARRS surface gravity          \\
22    & Bibcode                       & --    & Reference paper for spectral type    \\  
23    & Comment                & --    & Additional comment         \\ \bottomrule
\end{tabular}
\label{tab:sample}
\end{table*}

%% file: tables/SpecTypes.tex
\begin{table*}
\centering
\caption{Breakdown of the identified spectral types of the northern 40 pc sample.}
\begin{tabular}{@{}lll@{}}
\toprule
Spectral Type & Total Number & Model Composition \\ \midrule
DA            & 282          & pure-H (except He for two He-rich DA)\\
DAe           & 1            & pure-H\\
DAH or DAP    & 23        & pure-H\\
DAZ           & 21           & pure-H \\
DAZH          & 1            & pure-H \\
DB            & 1            & H/He=$10^{-5}$\\
DB+dM         & 1            & H/He=$10^{-5}$\\
DBA           & 2            & H/He=$10^{-5}$\\
DBAZ          & 1            & H/He=$10^{-5}$\\
DBP           & 1            & H/He=$10^{-5}$\\
DC            & 145          & H/He=$10^{-5}$, pure-He below 7000\,K, assumed pure-H below 5000\,K\\
DCP           & 2            & H/He=$10^{-5}$, pure-He below 7000\,K\\
DQ            & 14           & H/He=$10^{-5}$, pure-He below 7000\,K\\
DQP           & 1            & H/He=$10^{-5}$, pure-He below 7000\,K\\
DQpecP        & 1            & H/He=$10^{-5}$, pure-He below 7000\,K\\
DQZA          & 1            & H/He=$10^{-5}$, pure-He below 7000\,K\\
DZ            & 13           & H/He=$10^{-5}$, pure-He below 7000\,K\\
DZA           & 4            & H/He=$10^{-5}$, pure-He below 7000\,K (except H for one H-rich DZA)\\
DZH or DZP    & 3            & H/He=$10^{-5}$, pure-He below 7000\,K\\
DX or DXP     & 3            & assumed H/He=$10^{-5}$, pure-He below 7000\,K\\ 
Unknown       & 3            & assumed pure-H\\ \bottomrule
\end{tabular}
\label{tab:spect_table}
\end{table*}

%% file: tables/Unres_double_WD.tex
\begin{table*}
\centering
\caption{Double Degenerates in the Gaia DR2 Northern 40\,pc Sample}
\begin{tabular}{@{}lllllll@{}}
\toprule
WDJ name & Parallax [mas] & SpT & Gaia $\log g$$^{1}$ & Spectro $\log g$$^{1}$ & Orb. Period [day] &  Ref (Binarity) \\
\hline
\multicolumn{5}{l}{Double degenerate component of a triple or quadruple system} \\
\hline
J010349.92+050430.57 & 44.86 (0.12) & DA & 7.52 (0.01) & 8.17 (0.05) & 1.2 or 6.4 & \citet{maxted2000a}\\
J170530.44+480312.36 &  25.38 (0.03) & DA & 7.24 (0.01) & 7.67 (0.05) & 0.1448 & \citet{maxted2000b}\\
\hline
\multicolumn{5}{l}{Known double degenerate} \\
\hline
J053620.21+412955.62 & 30.99 (0.04) & DA & 7.23 (0.01) & 7.97 (0.05) &--  &\citet{zuckerman2003} \\
J094846.64+242125.88  & 27.37 (0.07) & DA& 8.29 (0.01) & 8.40 (0.11) &-- &\citet{liebert1993}\\
J131257.90+580511.29  & 31.32 (0.24) & DA& 7.64 (0.02) & 8.15 (0.05) &-- &\citet{gentile2018}\\
J164136.61+151237.93 & 31.48 (0.11) & DA & 7.87 (0.02) & 8.49 (0.07) & 1471 &\citet{harris03}\\
\hline
\multicolumn{5}{l}{Double degenerate candidate (also in literature)} \\
\hline
J012924.26+102301.34& 34.14 (0.05) & DA & 7.54 (0.01) & 7.88 (0.06) & -- &\citet{zuckerman2003} \\
J014511.23+313243.56 & 27.85 (0.09) & DA& 7.61 (0.02) & 8.12 (0.05) &  -- &\citet{bedard17}\\
J134532.97+420043.66  & 26.68 (0.07) & DC& 7.15 (0.04) & -- &  -- &\citet{limoges2015}\\
J163441.85+173634.09 & 39.05 (0.03) & DAZ & 7.26 (0.01) & 7.79 (0.05) & --  &\citet{toonen2017} \\
J205020.65+263040.76  & 52.34 (0.05) & DA& 7.27 (0.01) & -- & --  &\citet{hollands2018gaia}\\
J211345.93+262133.27 & 32.34 (0.32) & DA & 7.53 (0.02) & 8.15 (0.06) & --  &\citet{bergeron2001}\\
J225123.02+293944.49  & 51.47 (0.14) & DA& 7.71	(0.01) & -- & --  &\citet{hollands2018gaia}\\
J232519.87+140339.61 & 42.34 (0.13) & DA& 7.31 (0.01) & -- &  -- &\citet{limoges2015}\\
\hline
\multicolumn{5}{l}{Double degenerate candidate (this work)} \\
\hline
J000754.11+394732.18 & 29.03 (0.06) & DC & 6.88 (0.02) & -- &  -- &--\\
J002215.19+423642.15 & 29.32 (0.06) & DC& 7.58 (0.01) & -- & -- &--\\
J020847.22+251409.97 & 25.61 (0.05) & DA& 7.48 (0.01) & 7.91	(0.05) & -- & --\\
J023117.04+285939.88 & 38.47 (0.38) & DA& 7.67 (0.02) & -- & -- &--\\
J032020.30+233331.72 & 25.48 (0.19) & DC & 7.21 (0.05) & -- & -- &--\\
J054457.66+260300.14 & 27.68 (0.10) & DC& 7.11 (0.05) & & -- &--\\
J111536.96+003317.11 & 25.22 (0.14) & DA& 7.56 (0.04) & -- & -- &--\\
J192359.24+214103.62 & 28.73 (0.15) & DA& 7.55 (0.02) & 8.06 (0.02) & -- &--\\
J200654.88+614310.27 & 29.84 (0.09) & DA& 7.35 (0.02) & -- & -- &--\\
J214913.61+041550.35 & 28.44 (0.09) & DA& 7.40 (0.04) &-- &-- & --\\
\hline
\multicolumn{5}{l}{Low-mass white dwarf} \\
\hline
J094639.07+435452.24 & 31.29 (0.04) & DA & 7.69 (0.05) & 7.59 (0.01) & -- & \citet{brown2011}\\
J102459.83+044610.50 & 25.22 (0.49) & DA & 7.53 (0.04) & 7.63 (0.05) & 1.157  & \citet{brown2011} \\
\bottomrule
\end{tabular}
\label{tab:DoubleWD}\\
{Note: $^{1}$All photometric and spectroscopic fits are assuming a single white dwarf and are shown for illustrative purpose only. Spectroscopic parameters are from \citet{limoges2015} or \citet{tremblay2020_WHT}. Both studies account for 3D convective effects \citep{tremblay2013}. }
\end{table*}

%% file: tables/magnetic_objs.tex
\begin{table*}
\centering
\caption{Magnetic White Dwarfs in the Gaia DR2 Northern 40\,pc Sample}
\begin{tabular}{@{}llll@{}}
\toprule
WDJ name & SpT & $\langle B \rangle$& Ref \\
 &  & [MG] & \\
\hline
001214.75+502520.74	& DAH &	0.25	& \citet{landstreet2019}   \\
024208.44+111233.00	& DAH &	0.7	& \citet{ferrario2015} \\
025959.15+081156.43	& DAH &	0.1	& \citet{ferrario2015} \\
030350.56+060748.75	& DXP & 500	& \citet{landstreet2020}\\
033320.37+000720.65	& DAH &	850	& \citet{ferrario2015}\\
051553.54+283916.81	& DAH &	1.1	& \citet{limoges2015}\\
053714.90+675950.51	& DAH &	0.7	& \citet{tremblay2020_WHT}\\
063235.80+555903.12	& DAH &	1	& \citet{tremblay2020_WHT}\\
064400.61+092605.76	& DAH &	3.2	& \citet{tremblay2020_WHT}\\
064926.55+752124.97	& DAH & 	9	& \citet{tremblay2020_WHT}\\
073330.88+640927.44	& DAP &	0.1	& \citet{ferrario2015}\\
075959.58+433521.10	& DCP &	200	& \citet{ferrario2015}\\
084516.87+611704.81	& DAH &	0.8	& \citet{tremblay2020_WHT}\\
085830.87+412635.75	& DAH &	3.4	& \citet{ferrario2015}\\
091556.08+532523.86	& DCP &	100	& \citet{ferrario2015}\\
094846.64+242125.88	& DAP &	670	& \citet{ferrario2015}\\
101141.58+284559.07	& DQpecP &	100	& \citet{ferrario2015}\\
102907.46+112719.28	& DAH &	18	& \citet{ferrario2015}\\
123752.23+415624.69	& DQP & -- & \citet{ferrario2015}\\
130841.20+850228.16	& DAP &	4.9	& \citet{landstreet2019}   \\
133059.42+302953.65	& DZH &	0.7	& \citet{ferrario2015}\\
151534.80+823028.99	& DZH &	3	& \citet{tremblay2020_WHT}\\
153505.81+124745.20	& DZH &	0.3	& \citet{bagnulo19}\\
164057.15+534109.32	& DAH &	13	& \citet{ferrario2015}\\
165445.69+382936.63	& DAZH &	0.1 & \citet{ferrario2015}\\
165948.42+440104.04	& DAH &	2.3	& \citet{ferrario2015}\\
171450.80+391837.43	& DAH &	1.3	& \citet{ferrario2015}\\
174807.99+705235.92	& DXP &	100	& \citet{ferrario2015}\\
181608.87+245442.85	& DAP &	15	& \citet{ferrario2015}\\
183020.27+544727.21	& DXP &	170	& \citet{ferrario2015}\\
190010.25+703951.42	& DAP &	320	& \citet{landstreet2019}  \\ 
201222.27+311348.88	& DBP & 520	& \citet{ferrario2015}\\
204906.70+372814.05	& DAP & 0.06	& \citet{landstreet2019}   \\
215140.11+591734.85	& DAH & 0.8	& \citet{landstreet2019}   \\
233203.52+265846.12	& DAH & 2.3	& \citet{ferrario2015}\\
\bottomrule
\end{tabular}
\label{tab:magnetic}
\end{table*}

%% file: tables/UCWD.tex
\begin{table*}
\caption{Ultra-cool white dwarfs in the \gaia\ DR2 northern 40\,pc sample. Error bars should be interpreted with caution because of systematic uncertainties in predicted CIA opacities.}
\begin{tabular}{@{}llllll@{}}
\toprule
WDJ Name               & SpT    & \teff [K] & log(g) & Mass [M$_\odot$] & Ref                 \\ \midrule
J034646.52+245602.67 & DC     & 2970 $\pm$ 40 & 7.66  $\pm$ 0.30 & 0.39 $\pm$ 0.10 & \citet{limoges2015} \\
J110217.52+411321.18 & DC     & 3860 $\pm$ 30 & --  & -- & \citet{limoges2015} \\
J165401.26+625354.91 & DC     & 3080 $\pm$ 100 & 7.24  $\pm$ 0.03 & 0.22 $\pm$ 0.01 & \citet{limoges2015} \\
J140324.75+453333.02 & DC     & 2670 $\pm$ 1500 & --  & -- & \citet{kilic10}\\ \bottomrule
\end{tabular}
\label{tab:UCWD}
\end{table*}

%% file: tables/TABLEA1.tex
\begin{table*}
\centering
\caption{The catalogue of 524 \textit{Gaia} white dwarfs in the northern hemisphere and within 40\,pc can be accessed online at \href{http://deneb.astro.warwick.ac.uk/phrgwr/40pcTables/index.html}{this link}. See Table\,\ref{tab:sample} for content description.}
\label{tab:A1}
\end{table*}

%% file: tables/TABLEA2.tex
\begin{table*}
\centering
\caption{List of 64 main-sequence star contaminants and spectroscopically unobserved low probability white dwarf candidates within 40\,pc, in the northern hemisphere and drawn from the selection of \citet{gentile2018gaia}. The data can be accessed online at \href{http://deneb.astro.warwick.ac.uk/phrgwr/40pcTables/index.html}{this link}. See Table\,\ref{tab:sample} for content description.}
\label{tab:A2}
\end{table*}

%% file: tables/TABLEA4.tex
\begin{table*}
\caption{White dwarfs in the northern hemisphere that are likely within 40\,pc and missing from Table~\ref{tab:A1} and the input sample of \citet{gentile2018gaia}, sorting by increasing distance.}
\begin{tabular}{@{}lllllll@{}}
\toprule
Gaia DR2 ID & Name              &   Parallax [mas]          & Ref &  SpT  & Ref & Note \\ \midrule
\multicolumn{7}{l}{Confirmed 40\,pc Members} \\
\hline
--                   & WD 0736+053 &	284.56 $\pm$ 1.26	& (1) & DQZ& \citet{limoges2015} & (a)\\
3320184202856027776	 & WD 0553+053 &    125.0  $\pm$ 3.6    & (2) & DAH& \citet{limoges2015} & (b)\\
--                  & WD 1334+039 &    121.4 $\pm$ 3.4      & (2) & DA & \citet{limoges2015} & (c)\\
975968340912004352	 & WD 0727+482A&	88.543 $\pm$ 0.066  & (3) & DA & \citet{limoges2015} & (d)\\
975968340910692736	 & WD 0727+482B&    88.543 $\pm$ 0.066  & (3) & DA & \citet{limoges2015} & (d)\\
3978879594463069312	 & WD 1121+216 &    74.4 $\pm$ 2.8	    & (2) & DA & \citet{limoges2015} & (b)\\
1362295082910131200	& HD 159062   &  46.123 $\pm$ 0.024 & (3) & G9V+WD &  \citet{hirsch2019} & (e) \\
2274076301516712704  & WD 2126+734B & 	45.15 $\pm$ 0.21    & (4) & DC & \citet{zuckerman97} & (f)\\		
3701290326205270528	 & WD 1214+032 &    42.784 $\pm$ 0.063  & (4) & DA & \citet{limoges2015} & (f)\\
3817534337626005632	& WD 1120+073 &     31.23 $\pm$ 0.80    & (3) & DC & \citet{limoges2015} & (d)\\
3920187251456355072	& WD 1153+135 &     28.29 $\pm$ 0.66    & (5) & DC & \citet{leggett2018}& (b)\\
1962707287281651712	& PM J22105+4532 &  27.759 $\pm$ 0.088  & (4) & DC & \citet{limoges2015} & (f)\\
307323228064848512	 & WD 0108+277 &    26.35 $\pm$ 0.11    & (4) & DAZ& \citet{kawka2006} & (f)\\
\hline
\multicolumn{7}{l}{Unresolved Binaries} \\
\hline
1005873614079882880	& LHS 1817 & 61.13 $\pm$ 0.15 & (4) & M4.5V+WD & \citet{winters2020} & (g) \\
1548104507825815296	& WD 1213+528 &  34.834 $\pm$ 0.032 & (4) & DA+dM & \citet{limoges2015} & (g) \\
1550299304833675392	& WD 1324+458 &  32.734 $\pm$ 0.030	& (4) & M3V+DA & \citet{parsons2010} & (g) \\
3845263368043086080	& WD 0911+023 &  28.40 $\pm$ 0.37 & (4)  & B9.5V+WD & \citet{holberg2013} & (g) \\
4478524169500496000	& HD 169889 &  28.30 $\pm$ 0.07 & (4)  & G9V+WD & \citet{crepp2018} & (g) \\
3831059120921201280	& WD 1026+002 &  25.064 $\pm$ 0.060 & (4) & DA+dM & \citet{koester09} & (g) \\
\hline
\multicolumn{7}{l}{Possible 40\,pc Members} \\
\hline
4018536882933053056	& WD 1132+275 & --       & -- & DC & \citet{limoges2015} & (b)\\	
1267487150183614976 & WD 1143+256 & -- 	      & -- & DA & \citet{limoges2015} & (h) \\
2701893698904233216	& WD 2140+078 & --       & -- & DA & \citet{limoges2015} & (b)\\
 \bottomrule

\label{tab:A4}
\end{tabular}\\
\justifying
    {\noindent}References: (1) \citet{vanLeeuwen07}, (2) \citet{vanAltena95}, (3) \textit{Gaia} DR2 (companion),  (4) \textit{Gaia} DR2 (white dwarf), (5) \citet{leggett2018}.\\ Notes: (a) No \textit{Gaia} detection due to saturation of Procyon A. (b) White dwarf does not have DR2 five-parameter astrometry. (c) No \textit{Gaia} detection, noting that the white dwarf has an $\approx$ 4$^{\prime\prime}$ yr$^{-1}$ proper motion. (d) White dwarf does not have DR2 five-parameter astrometry, but known companion does. (e) White dwarf resolved at 2.$^{\prime\prime}$7 according to \citet{hirsch2019} but no \textit{Gaia} DR2 source detected corresponding to the white dwarf. (f) DR2 five-parameter astrometry available but white dwarf absent from \citet{gentile2018gaia} because of missing or incorrect colours (large BP/RP excess factor). (g) Missing from \citet{gentile2018gaia} because of the important flux contribution from the non-degenerate companion in the optical. (h) Distance estimate of $<$ 40\,pc in \citet{limoges2015} based on atmospheric parameters. Detected with the established proper motion, but with a parallax of 1.44 $\pm$ 0.55 mas. 
\end{table*}

%% file: tables/TABLEA3.tex
\begin{table*}
\caption{\textit{Gaia} white dwarf candidates in the northern hemisphere that may lie within 40\,pc based on \textit{Gaia} parallax errors. The upper rows in each section are the objects within 1$\sigma$, and the lower those which lie within 2$\sigma$.}
\begin{tabular}{@{}llllll@{}}
\toprule
WDJ Name               & Gaia DR2 ID & WD Name &  Parallax [mas]          & SpT        & Ref       \\ \midrule
\multicolumn{5}{l}{Confirmed White Dwarfs} \\
\hline
WDJ001339.15+001924.58 & 2545505281002947200 & WD 0011+000 & 24.96 (0.06)    & DA   & \citet{gianninas2011}\\
WDJ055231.03+164250.27 & 3349849778193723008 & --          & 24.97 (0.06)    & DBA  & \citet{tremblay2020_WHT}\\
WDJ080247.02+564640.62 & 1081514379072280320 & --          & 24.87 (0.21)    & DC   & \citet{tremblay2020_WHT}\\
WDJ102203.66+824310.00 & 1146403741412820864 & --          & 24.95 (0.13)    & DA   & \citet{tremblay2020_WHT}\\
WDJ134118.69+022737.01 & 3713218786120541824 & WD 1338+027 & 24.88 (0.13)    & DQ   & \citet{kilic10}\\
WDJ171430.49+212710.45 & 4567158653660872064 & WD 1712+215 & 24.96 (0.06)    & DC   & \citet{putney97}\\
WDJ180218.60+135405.46 & 4496751667093478016 & --          & 24.97 (0.07)    & DAZ  & \citet{tremblay2020_WHT}\\
WDJ184733.18+282057.54 & 4539227892919675648 & --          & 24.87 (0.16)    & DC   & \citet{tremblay2020_WHT}\\
WDJ192206.20+023313.29 & 4288942973032203904 & --          & 24.95 (0.32)    & DZ   & \citet{tremblay2020_WHT}\\
WDJ202956.18+391332.20 & 2064272612307218176 & WD 2028+390 & 24.96 (0.04)    & DA   & \citet{gianninas2011}\\
WDJ221321.31+034911.08 & 2707796667595813248 & NLTT 53229  & 24.68 (0.33)    & DC   & \citet{limoges2015}\\
WDJ235750.73+194905.90 & 2822330113802737408 & --          & 24.92 (0.12)    & DZ   & \citet{tremblay2020_WHT}\\
&&&\\
WDJ070845.80+204451.70 & 3366672379112835328 & PM J07087+2044 & 24.80 (0.10)    & DA   & \citet{limoges2015}\\
WDJ112105.81+375615.39 & 763981296484951936  & CBS 429        & 24.91 (0.06)    & DA   & \citet{limoges2015}\\
WDJ133359.84+001655.03 & 3662951038644235776 & WD 1331+005    & 24.45 (0.35)    & DQ   & \citet{ferrario2015}\\
WDJ134043.36+020348.30 & 3663664003222454528 & WD 1338+023    & 24.69 (0.16)    & DC   & \citet{leggett2018}\\
WDJ231845.10+123602.77 & 2811321837744375936 & WD 2316+123    & 24.87 (0.08)    & DAH  & \citet{limoges2015}\\
\hline
\multicolumn{5}{l}{Main Sequence Contaminants} \\
\hline
WDJ005645.62+551556.10 & 423445945315773440  &  -- & 24.36 (0.89)    & Star   & \citet{tremblay2020_WHT}\\
WDJ134252.41+003312.28 & 3663164069021692800 & -- & 24.19 (0.96)    & Star   & \citet{tremblay2020_WHT}\\
\hline
\multicolumn{5}{l}{Unobserved Objects} \\
\hline
WDJ015348.12+654946.61 & 518443341936881664  & --     & 22.90 (2.49)    & --     & Low $P_{\rm WD}$\\
WDJ171605.95+190544.28 & 4548611473051368576 & --     & 24.43 (0.90)    & --     & Low $P_{\rm WD}$\\
WDJ181131.97+132601.58 & 4497270567861853440 & --     & 24.52 (0.50)    & --     & Low $P_{\rm WD}$\\
WDJ194052.95+170459.04 & 1823962663799965440 & --     & 24.82 (0.97)    & --     & Low $P_{\rm WD}$\\
WDJ194943.60+152641.56 & 1819762318874306048 & --     & 24.53 (0.55)    & --     & Low $P_{\rm WD}$\\
&&&\\
WDJ002955.72+472645.48 & 389482855766584704  & --     & 22.91 (1.30)    & --     & -- \\
WDJ050647.89+203014.55 & 3408829849653432576 & LP 416-350     & 24.76 (0.15)    & --   & --\\
WDJ055326.34+062759.60 & 3322605013926459520 & --             & 24.30 (0.54)    & --   & --\\
WDJ112542.42+041318.02 & 3812805230740561280 & --     & 21.88 (2.10)    & --     & Low $P_{\rm WD}$\\
WDJ115016.52+154700.09 & 3924562444445708288 & --     & 21.81 (2.44)    & --     & Low $P_{\rm WD}$\\
WDJ122048.91+482912.98 & 1545564017495017088 & --     & 22.91 (1.29)    & --     & Low $P_{\rm WD}$\\
WDJ123010.77+100537.85 & 3903961547910932992 & --     & 22.00 (2.59)    & --     & Low $P_{\rm WD}$\\
WDJ174129.76+015632.44 & 4375912002010905600 & --     & 23.77 (0.93)    & --   & Low $P_{\rm WD}$\\
WDJ194553.70+180829.53 & 1824144048862367104 & --     & 23.26 (1.05)    & --     & Low $P_{\rm WD}$\\ \bottomrule
\end{tabular}
\label{tab:A3}
\end{table*}

%% file: tables/TABLEA5.tex
\begin{table*}
\centering
\caption{The catalogue of 56 wide binary systems including a white dwarf in the northern hemisphere and within 40\,pc can be accessed online at \href{http://deneb.astro.warwick.ac.uk/phrgwr/40pcTables/index.html}{this link}.}
\label{tab:A5}
\justifying
    {\noindent}Note: (1) Found in \citet{el-badry2018binary}, (2) Found in \citet{hollands2018gaia}.
\end{table*}

%% file: wd40pc.bbl
\begin{thebibliography}{}
\makeatletter
\relax
\def\mn@urlcharsother{\let\do\@makeother \do\$\do\&\do\#\do\^\do\_\do\%\do\~}
\def\mn@doi{\begingroup\mn@urlcharsother \@ifnextchar [ {\mn@doi@}
  {\mn@doi@[]}}
\def\mn@doi@[#1]#2{\def\@tempa{#1}\ifx\@tempa\@empty \href
  {http://dx.doi.org/#2} {doi:#2}\else \href {http://dx.doi.org/#2} {#1}\fi
  \endgroup}
\def\mn@eprint#1#2{\mn@eprint@#1:#2::\@nil}
\def\mn@eprint@arXiv#1{\href {http://arxiv.org/abs/#1} {{\tt arXiv:#1}}}
\def\mn@eprint@dblp#1{\href {http://dblp.uni-trier.de/rec/bibtex/#1.xml}
  {dblp:#1}}
\def\mn@eprint@#1:#2:#3:#4\@nil{\def\@tempa {#1}\def\@tempb {#2}\def\@tempc
  {#3}\ifx \@tempc \@empty \let \@tempc \@tempb \let \@tempb \@tempa \fi \ifx
  \@tempb \@empty \def\@tempb {arXiv}\fi \@ifundefined
  {mn@eprint@\@tempb}{\@tempb:\@tempc}{\expandafter \expandafter \csname
  mn@eprint@\@tempb\endcsname \expandafter{\@tempc}}}

\bibitem[\protect\citeauthoryear{{Bagnulo} \& {Landstreet}}{{Bagnulo} \&
  {Landstreet}}{2019}]{bagnulo19}
{Bagnulo} S.,  {Landstreet} J.~D.,  2019, \mn@doi [\aap]
  {10.1051/0004-6361/201936068}, \href
  {https://ui.adsabs.harvard.edu/abs/2019A&A...630A..65B} {630, A65}

\bibitem[\protect\citeauthoryear{{Becklin} \& {Zuckerman}}{{Becklin} \&
  {Zuckerman}}{1988}]{becklin88}
{Becklin} E.~E.,  {Zuckerman} B.,  1988, \mn@doi [\nat] {10.1038/336656a0},
  \href {https://ui.adsabs.harvard.edu/abs/1988Natur.336..656B} {336, 656}

\bibitem[\protect\citeauthoryear{{B{\'e}dard}, {Bergeron}  \&
  {Fontaine}}{{B{\'e}dard} et~al.}{2017}]{bedard17}
{B{\'e}dard} A.,  {Bergeron} P.,   {Fontaine} G.,  2017, \mn@doi [\apj]
  {10.3847/1538-4357/aa8bb6}, \href
  {https://ui.adsabs.harvard.edu/abs/2017ApJ...848...11B} {848, 11}

\bibitem[\protect\citeauthoryear{{Bergeron}, {Ruiz}  \& {Leggett}}{{Bergeron}
  et~al.}{1997}]{bergeron1997}
{Bergeron} P.,  {Ruiz} M.~T.,   {Leggett} S.~K.,  1997, \mn@doi [\apjs]
  {10.1086/312955}, \href
  {https://ui.adsabs.harvard.edu/abs/1997ApJS..108..339B} {108, 339}

\bibitem[\protect\citeauthoryear{{Bergeron}, {Leggett}  \& {Ruiz}}{{Bergeron}
  et~al.}{2001}]{bergeron2001}
{Bergeron} P.,  {Leggett} S.~K.,   {Ruiz} M.~T.,  2001, \mn@doi [\apjs]
  {10.1086/320356}, \href
  {https://ui.adsabs.harvard.edu/abs/2001ApJS..133..413B} {133, 413}

\bibitem[\protect\citeauthoryear{{Bergeron} et~al.,}{{Bergeron}
  et~al.}{2011}]{heliummodelbergeron2011}
{Bergeron} P.,  et~al., 2011, \mn@doi [\apj] {10.1088/0004-637X/737/1/28},
  \href {https://ui.adsabs.harvard.edu/abs/2011ApJ...737...28B} {737, 28}

\bibitem[\protect\citeauthoryear{{Bergeron}, {Dufour}, {Fontaine}, {Coutu},
  {Blouin}, {Genest-Beaulieu}, {B{\'e}dard}  \& {Rolland }}{{Bergeron}
  et~al.}{2019}]{bergeron2019}
{Bergeron} P.,  {Dufour} P.,  {Fontaine} G.,  {Coutu} S.,  {Blouin} S.,
  {Genest-Beaulieu} C.,  {B{\'e}dard} A.,   {Rolland } B.,  2019, \mn@doi
  [\apj] {10.3847/1538-4357/ab153a}, \href
  {https://ui.adsabs.harvard.edu/abs/2019ApJ...876...67B} {876, 67}

\bibitem[\protect\citeauthoryear{{Blouin}, {Kowalski}  \& {Dufour}}{{Blouin}
  et~al.}{2017}]{blouin2017}
{Blouin} S.,  {Kowalski} P.~M.,   {Dufour} P.,  2017, \mn@doi [\apj]
  {10.3847/1538-4357/aa8ad6}, \href
  {https://ui.adsabs.harvard.edu/abs/2017ApJ...848...36B} {848, 36}

\bibitem[\protect\citeauthoryear{{Blouin}, {Dufour}, {Thibeault}  \&
  {Allard}}{{Blouin} et~al.}{2019}]{blouin2019}
{Blouin} S.,  {Dufour} P.,  {Thibeault} C.,   {Allard} N.~F.,  2019, \mn@doi
  [\apj] {10.3847/1538-4357/ab1f82}, \href
  {https://ui.adsabs.harvard.edu/abs/2019ApJ...878...63B} {878, 63}

\bibitem[\protect\citeauthoryear{{Brown}, {Kilic}, {Brown}  \&
  {Kenyon}}{{Brown} et~al.}{2011}]{brown2011}
{Brown} J.~M.,  {Kilic} M.,  {Brown} W.~R.,   {Kenyon} S.~J.,  2011, \mn@doi
  [\apj] {10.1088/0004-637X/730/2/67}, \href
  {https://ui.adsabs.harvard.edu/abs/2011ApJ...730...67B} {730, 67}

\bibitem[\protect\citeauthoryear{{Chambers} et~al.,}{{Chambers}
  et~al.}{2016}]{panstarrs}
{Chambers} K.~C.,  et~al., 2016, arXiv e-prints, \href
  {https://ui.adsabs.harvard.edu/abs/2016arXiv161205560C} {p. arXiv:1612.05560}

\bibitem[\protect\citeauthoryear{{Chen} \& {Hansen}}{{Chen} \&
  {Hansen}}{2012}]{chen12}
{Chen} E.~Y.,  {Hansen} B. M.~S.,  2012, \mn@doi [\apjl]
  {10.1088/2041-8205/753/1/L16}, \href
  {https://ui.adsabs.harvard.edu/abs/2012ApJ...753L..16C} {753, L16}

\bibitem[\protect\citeauthoryear{{Cheng}, {Cummings}  \& {M{\'e}nard}}{{Cheng}
  et~al.}{2019}]{cheng2019}
{Cheng} S.,  {Cummings} J.~D.,   {M{\'e}nard} B.,  2019, \mn@doi [\apj]
  {10.3847/1538-4357/ab4989}, \href
  {https://ui.adsabs.harvard.edu/abs/2019ApJ...886..100C} {886, 100}

\bibitem[\protect\citeauthoryear{{Coutu}, {Dufour}, {Bergeron}, {Blouin},
  {Loranger}, {Allard}  \& {Dunlap}}{{Coutu} et~al.}{2019}]{coutu2019}
{Coutu} S.,  {Dufour} P.,  {Bergeron} P.,  {Blouin} S.,  {Loranger} E.,
  {Allard} N.~F.,   {Dunlap} B.~H.,  2019, \mn@doi [\apj]
  {10.3847/1538-4357/ab46b9}, \href
  {https://ui.adsabs.harvard.edu/abs/2019ApJ...885...74C} {885, 74}

\bibitem[\protect\citeauthoryear{{Crepp} et~al.,}{{Crepp}
  et~al.}{2018}]{crepp2018}
{Crepp} J.~R.,  et~al., 2018, \mn@doi [\apj] {10.3847/1538-4357/aad381}, \href
  {https://ui.adsabs.harvard.edu/abs/2018ApJ...864...42C} {864, 42}

\bibitem[\protect\citeauthoryear{{Cunningham}, {Tremblay}, {Gentile Fusillo},
  {Hollands}  \& {Cukanovaite}}{{Cunningham} et~al.}{2020}]{cunningham2020}
{Cunningham} T.,  {Tremblay} P.-E.,  {Gentile Fusillo} N.~P.,  {Hollands} M.,
  {Cukanovaite} E.,  2020, \mn@doi [\mnras] {10.1093/mnras/stz3638}, \href
  {https://ui.adsabs.harvard.edu/abs/2020MNRAS.492.3540C} {492, 3540}

\bibitem[\protect\citeauthoryear{{Dufour}, {Blouin}, {Coutu},
  {Fortin-Archambault}, {Thibeault}, {Bergeron}  \& {Fontaine}}{{Dufour}
  et~al.}{2017}]{dufourMWDD}
{Dufour} P.,  {Blouin} S.,  {Coutu} S.,  {Fortin-Archambault} M.,  {Thibeault}
  C.,  {Bergeron} P.,   {Fontaine} G.,  2017, in {Tremblay} P.~E.,  {Gaensicke}
  B.,   {Marsh} T.,  eds,  Astronomical Society of the Pacific Conference
  Series Vol. 509, 20th European White Dwarf Workshop. p.~3

\bibitem[\protect\citeauthoryear{{El-Badry} \& {Rix}}{{El-Badry} \&
  {Rix}}{2018}]{el-badry2018binary}
{El-Badry} K.,  {Rix} H.-W.,  2018, \mn@doi [\mnras] {10.1093/mnras/sty2186},
  \href {https://ui.adsabs.harvard.edu/abs/2018MNRAS.480.4884E} {480, 4884}

\bibitem[\protect\citeauthoryear{{El-Badry}, {Rix}  \& {Weisz}}{{El-Badry}
  et~al.}{2018}]{el-badry2018}
{El-Badry} K.,  {Rix} H.-W.,   {Weisz} D.~R.,  2018, \mn@doi [\apjl]
  {10.3847/2041-8213/aaca9c}, \href
  {https://ui.adsabs.harvard.edu/abs/2018ApJ...860L..17E} {860, L17}

\bibitem[\protect\citeauthoryear{{Fantin} et~al.,}{{Fantin}
  et~al.}{2019}]{fantin2019}
{Fantin} N.~J.,  et~al., 2019, \mn@doi [\apj] {10.3847/1538-4357/ab5521}, \href
  {https://ui.adsabs.harvard.edu/abs/2019ApJ...887..148F} {887, 148}

\bibitem[\protect\citeauthoryear{{Ferrario}, {de Martino}  \&
  {G{\"a}nsicke}}{{Ferrario} et~al.}{2015}]{ferrario2015}
{Ferrario} L.,  {de Martino} D.,   {G{\"a}nsicke} B.~T.,  2015, \mn@doi [\ssr]
  {10.1007/s11214-015-0152-0}, \href
  {https://ui.adsabs.harvard.edu/abs/2015SSRv..191..111F} {191, 111}

\bibitem[\protect\citeauthoryear{{Fontaine}, {Brassard}  \&
  {Bergeron}}{{Fontaine} et~al.}{2001}]{fontaine2001}
{Fontaine} G.,  {Brassard} P.,   {Bergeron} P.,  2001, \mn@doi [\pasp]
  {10.1086/319535}, \href
  {https://ui.adsabs.harvard.edu/abs/2001PASP..113..409F} {113, 409}

\bibitem[\protect\citeauthoryear{{Fontaine}, {Brassard}, {Charpinet}, {Rand
  all}  \& {Van Grootel}}{{Fontaine} et~al.}{2013}]{fontaine13}
{Fontaine} G.,  {Brassard} P.,  {Charpinet} S.,  {Rand all} S.~K.,   {Van
  Grootel} V.,  2013, in European Physical Journal Web of Conferences. p.
  05001, \mn@doi{10.1051/epjconf/20134305001}

\bibitem[\protect\citeauthoryear{{Fuchs} \& {Jahrei{\ss}}}{{Fuchs} \&
  {Jahrei{\ss}}}{1998}]{fuchs199820pchalowd}
{Fuchs} B.,  {Jahrei{\ss}} H.,  1998, \aap, \href
  {https://ui.adsabs.harvard.edu/abs/1998A&A...329...81F} {329, 81}

\bibitem[\protect\citeauthoryear{{Gaia Collaboration} et~al.,}{{Gaia
  Collaboration} et~al.}{2018a}]{gaiadr2}
{Gaia Collaboration} et~al., 2018a, \mn@doi [\aap]
  {10.1051/0004-6361/201833051}, \href
  {https://ui.adsabs.harvard.edu/abs/2018A&A...616A...1G} {616, A1}

\bibitem[\protect\citeauthoryear{{Gaia Collaboration} et~al.,}{{Gaia
  Collaboration} et~al.}{2018b}]{gaiaHR}
{Gaia Collaboration} et~al., 2018b, \mn@doi [\aap]
  {10.1051/0004-6361/201832843}, \href
  {https://ui.adsabs.harvard.edu/abs/2018A&A...616A..10G} {616, A10}

\bibitem[\protect\citeauthoryear{{Genest-Beaulieu} \&
  {Bergeron}}{{Genest-Beaulieu} \& {Bergeron}}{2019}]{genest-beaulieu2019}
{Genest-Beaulieu} C.,  {Bergeron} P.,  2019, \mn@doi [\apj]
  {10.3847/1538-4357/aafac6}, \href
  {https://ui.adsabs.harvard.edu/abs/2019ApJ...871..169G} {871, 169}

\bibitem[\protect\citeauthoryear{{Gentile Fusillo}, {G{\"a}nsicke}, {Farihi},
  {Koester}, {Schreiber}  \& {Pala}}{{Gentile Fusillo} et~al.}{2017}]{ngf2017}
{Gentile Fusillo} N.~P.,  {G{\"a}nsicke} B.~T.,  {Farihi} J.,  {Koester} D.,
  {Schreiber} M.~R.,   {Pala} A.~F.,  2017, \mn@doi [\mnras]
  {10.1093/mnras/stx468}, \href
  {https://ui.adsabs.harvard.edu/abs/2017MNRAS.468..971G} {468, 971}

\bibitem[\protect\citeauthoryear{{Gentile Fusillo}, {Tremblay}, {Jordan},
  {G{\"a}nsicke}, {Kalirai}  \& {Cummings}}{{Gentile Fusillo}
  et~al.}{2018}]{gentile2018}
{Gentile Fusillo} N.~P.,  {Tremblay} P.~E.,  {Jordan} S.,  {G{\"a}nsicke}
  B.~T.,  {Kalirai} J.~S.,   {Cummings} J.,  2018, \mn@doi [\mnras]
  {10.1093/mnras/stx2584}, \href
  {https://ui.adsabs.harvard.edu/abs/2018MNRAS.473.3693G} {473, 3693}

\bibitem[\protect\citeauthoryear{{Gentile Fusillo} et~al.,}{{Gentile Fusillo}
  et~al.}{2019}]{gentile2018gaia}
{Gentile Fusillo} N.~P.,  et~al., 2019, \mn@doi [\mnras]
  {10.1093/mnras/sty3016}, \href
  {https://ui.adsabs.harvard.edu/abs/2019MNRAS.482.4570G} {482, 4570}

\bibitem[\protect\citeauthoryear{{Gentile Fusillo}, {Tremblay}, {Bohlin},
  {Deustua}  \& {Kalirai}}{{Gentile Fusillo} et~al.}{2020}]{gentile2020}
{Gentile Fusillo} N.~P.,  {Tremblay} P.-E.,  {Bohlin} R.~C.,  {Deustua} S.~E.,
   {Kalirai} J.~S.,  2020, \mn@doi [\mnras] {10.1093/mnras/stz2984}, \href
  {https://ui.adsabs.harvard.edu/abs/2020MNRAS.491.3613G} {491, 3613}

\bibitem[\protect\citeauthoryear{{Giammichele}, {Bergeron}  \&
  {Dufour}}{{Giammichele} et~al.}{2012}]{giammichele2012know}
{Giammichele} N.,  {Bergeron} P.,   {Dufour} P.,  2012, \mn@doi [\apjs]
  {10.1088/0067-0049/199/2/29}, \href
  {https://ui.adsabs.harvard.edu/abs/2012ApJS..199...29G} {199, 29}

\bibitem[\protect\citeauthoryear{{Gianninas}, {Bergeron}  \&
  {Ruiz}}{{Gianninas} et~al.}{2011}]{gianninas2011}
{Gianninas} A.,  {Bergeron} P.,   {Ruiz} M.~T.,  2011, \mn@doi [\apj]
  {10.1088/0004-637X/743/2/138}, \href
  {https://ui.adsabs.harvard.edu/abs/2011ApJ...743..138G} {743, 138}

\bibitem[\protect\citeauthoryear{{Hall}, {Kowalski}, {Harris}, {Awal},
  {Leggett}, {Kilic}, {Anderson}  \& {Gates}}{{Hall}
  et~al.}{2008}]{Hall2008halowd}
{Hall} P.~B.,  {Kowalski} P.~M.,  {Harris} H.~C.,  {Awal} A.,  {Leggett} S.~K.,
   {Kilic} M.,  {Anderson} S.~F.,   {Gates} E.,  2008, \mn@doi [\aj]
  {10.1088/0004-6256/136/1/76}, \href
  {https://ui.adsabs.harvard.edu/abs/2008AJ....136...76H} {136, 76}

\bibitem[\protect\citeauthoryear{{Hambly}, {Smartt}  \& {Hodgkin}}{{Hambly}
  et~al.}{1997}]{hambly1997halowd}
{Hambly} N.~C.,  {Smartt} S.~J.,   {Hodgkin} S.~T.,  1997, \mn@doi [\apjl]
  {10.1086/316797}, \href
  {https://ui.adsabs.harvard.edu/abs/1997ApJ...489L.157H} {489, L157}

\bibitem[\protect\citeauthoryear{{Harris} et~al.,}{{Harris}
  et~al.}{2013}]{harris03}
{Harris} H.~C.,  et~al., 2013, \mn@doi [\apj] {10.1088/0004-637X/779/1/21},
  \href {https://ui.adsabs.harvard.edu/abs/2013ApJ...779...21H} {779, 21}

\bibitem[\protect\citeauthoryear{{Haywood}, {Di Matteo}, {Lehnert}, {Snaith},
  {Khoperskov}  \& {G{\'o}mez}}{{Haywood} et~al.}{2018}]{haywood2018}
{Haywood} M.,  {Di Matteo} P.,  {Lehnert} M.~D.,  {Snaith} O.,  {Khoperskov}
  S.,   {G{\'o}mez} A.,  2018, \mn@doi [\apj] {10.3847/1538-4357/aad235}, \href
  {https://ui.adsabs.harvard.edu/abs/2018ApJ...863..113H} {863, 113}

\bibitem[\protect\citeauthoryear{{Helmi}, {Babusiaux}, {Koppelman}, {Massari},
  {Veljanoski}  \& {Brown}}{{Helmi} et~al.}{2018}]{helmi2018}
{Helmi} A.,  {Babusiaux} C.,  {Koppelman} H.~H.,  {Massari} D.,  {Veljanoski}
  J.,   {Brown} A. G.~A.,  2018, \mn@doi [\nat] {10.1038/s41586-018-0625-x},
  \href {https://ui.adsabs.harvard.edu/abs/2018Natur.563...85H} {563, 85}

\bibitem[\protect\citeauthoryear{{Henry} et~al.,}{{Henry}
  et~al.}{2018}]{henry2018}
{Henry} T.~J.,  et~al., 2018, \mn@doi [\aj] {10.3847/1538-3881/aac262}, \href
  {https://ui.adsabs.harvard.edu/abs/2018AJ....155..265H} {155, 265}

\bibitem[\protect\citeauthoryear{{Hirsch} et~al.,}{{Hirsch}
  et~al.}{2019}]{hirsch2019}
{Hirsch} L.~A.,  et~al., 2019, \mn@doi [\apj] {10.3847/1538-4357/ab1b11}, \href
  {https://ui.adsabs.harvard.edu/abs/2019ApJ...878...50H} {878, 50}

\bibitem[\protect\citeauthoryear{{Holberg}, {Oswalt}  \& {Sion}}{{Holberg}
  et~al.}{2002}]{holberg2002}
{Holberg} J.~B.,  {Oswalt} T.~D.,   {Sion} E.~M.,  2002, \mn@doi [\apj]
  {10.1086/339842}, \href
  {https://ui.adsabs.harvard.edu/abs/2002ApJ...571..512H} {571, 512}

\bibitem[\protect\citeauthoryear{{Holberg}, {Sion}, {Oswalt}, {McCook}, {Foran}
   \& {Subasavage}}{{Holberg} et~al.}{2008}]{holberg2008}
{Holberg} J.~B.,  {Sion} E.~M.,  {Oswalt} T.,  {McCook} G.~P.,  {Foran} S.,
  {Subasavage} J.~P.,  2008, \mn@doi [\aj] {10.1088/0004-6256/135/4/1225},
  \href {https://ui.adsabs.harvard.edu/abs/2008AJ....135.1225H} {135, 1225}

\bibitem[\protect\citeauthoryear{{Holberg}, {Oswalt}, {Sion}, {Barstow}  \&
  {Burleigh}}{{Holberg} et~al.}{2013}]{holberg2013}
{Holberg} J.~B.,  {Oswalt} T.~D.,  {Sion} E.~M.,  {Barstow} M.~A.,   {Burleigh}
  M.~R.,  2013, \mn@doi [\mnras] {10.1093/mnras/stt1433}, \href
  {https://ui.adsabs.harvard.edu/abs/2013MNRAS.435.2077H} {435, 2077}

\bibitem[\protect\citeauthoryear{{Holberg}, {Oswalt}, {Sion}  \&
  {McCook}}{{Holberg} et~al.}{2016}]{holberg201625}
{Holberg} J.~B.,  {Oswalt} T.~D.,  {Sion} E.~M.,   {McCook} G.~P.,  2016,
  \mn@doi [\mnras] {10.1093/mnras/stw1357}, \href
  {https://ui.adsabs.harvard.edu/abs/2016MNRAS.462.2295H} {462, 2295}

\bibitem[\protect\citeauthoryear{{Hollands}, {G{\"a}nsicke}  \&
  {Koester}}{{Hollands} et~al.}{2018a}]{hollands2018planet}
{Hollands} M.~A.,  {G{\"a}nsicke} B.~T.,   {Koester} D.,  2018a, \mn@doi
  [\mnras] {10.1093/mnras/sty592}, \href
  {https://ui.adsabs.harvard.edu/abs/2018MNRAS.477...93H} {477, 93}

\bibitem[\protect\citeauthoryear{{Hollands}, {Tremblay}, {G{\"a}nsicke},
  {Gentile-Fusillo}  \& {Toonen}}{{Hollands} et~al.}{2018b}]{hollands2018gaia}
{Hollands} M.~A.,  {Tremblay} P.~E.,  {G{\"a}nsicke} B.~T.,  {Gentile-Fusillo}
  N.~P.,   {Toonen} S.,  2018b, \mn@doi [\mnras] {10.1093/mnras/sty2057}, \href
  {https://ui.adsabs.harvard.edu/abs/2018MNRAS.480.3942H} {480, 3942}

\bibitem[\protect\citeauthoryear{{Isern}}{{Isern}}{2019}]{isern2019}
{Isern} J.,  2019, \mn@doi [\apjl] {10.3847/2041-8213/ab238e}, \href
  {https://ui.adsabs.harvard.edu/abs/2019ApJ...878L..11I} {878, L11}

\bibitem[\protect\citeauthoryear{{Jim{\'e}nez-Esteban}, {Torres},
  {Rebassa-Mansergas}, {Skorobogatov}, {Solano}, {Cantero}  \&
  {Rodrigo}}{{Jim{\'e}nez-Esteban} et~al.}{2018}]{jimenez2018}
{Jim{\'e}nez-Esteban} F.~M.,  {Torres} S.,  {Rebassa-Mansergas} A.,
  {Skorobogatov} G.,  {Solano} E.,  {Cantero} C.,   {Rodrigo} C.,  2018,
  \mn@doi [\mnras] {10.1093/mnras/sty2120}, \href
  {https://ui.adsabs.harvard.edu/abs/2018MNRAS.480.4505J} {480, 4505}

\bibitem[\protect\citeauthoryear{{Kawka}}{{Kawka}}{2020}]{kawka2020}
{Kawka} A.,  2020, arXiv e-prints, \href
  {https://ui.adsabs.harvard.edu/abs/2020arXiv200110672K} {p. arXiv:2001.10672}

\bibitem[\protect\citeauthoryear{{Kawka} \& {Vennes}}{{Kawka} \&
  {Vennes}}{2006}]{kawka2006}
{Kawka} A.,  {Vennes} S.,  2006, \mn@doi [\apj] {10.1086/501451}, \href
  {https://ui.adsabs.harvard.edu/abs/2006ApJ...643..402K} {643, 402}

\bibitem[\protect\citeauthoryear{{Kawka} \& {Vennes}}{{Kawka} \&
  {Vennes}}{2012}]{kawka2012}
{Kawka} A.,  {Vennes} S.,  2012, \mn@doi [\mnras]
  {10.1111/j.1365-2966.2012.21574.x}, \href
  {https://ui.adsabs.harvard.edu/abs/2012MNRAS.425.1394K} {425, 1394}

\bibitem[\protect\citeauthoryear{{Kawka}, {Simpson}, {Vennes}, {Bessell}, {Da
  Costa}, {Marino}  \& {Murphy}}{{Kawka} et~al.}{2020}]{kawka2020b}
{Kawka} A.,  {Simpson} J.~D.,  {Vennes} S.,  {Bessell} M.~S.,  {Da Costa}
  G.~S.,  {Marino} A.~F.,   {Murphy} S.~J.,  2020, arXiv e-prints, \href
  {https://ui.adsabs.harvard.edu/abs/2020arXiv200407556K} {p. arXiv:2004.07556}

\bibitem[\protect\citeauthoryear{{Kilic} et~al.,}{{Kilic}
  et~al.}{2010}]{kilic10}
{Kilic} M.,  et~al., 2010, \mn@doi [\apjs] {10.1088/0067-0049/190/1/77}, \href
  {https://ui.adsabs.harvard.edu/abs/2010ApJS..190...77K} {190, 77}

\bibitem[\protect\citeauthoryear{{Kilic}, {Bergeron}, {Dame}, {Hambly},
  {Rowell}  \& {Crawford}}{{Kilic} et~al.}{2019}]{kilic2019}
{Kilic} M.,  {Bergeron} P.,  {Dame} K.,  {Hambly} N.~C.,  {Rowell} N.,
  {Crawford} C.~L.,  2019, \mn@doi [\mnras] {10.1093/mnras/sty2755}, \href
  {https://ui.adsabs.harvard.edu/abs/2019MNRAS.482..965K} {482, 965}

\bibitem[\protect\citeauthoryear{{Kilic}, {Bergeron}, {Kosakowski}, {Brown},
  {Agueros}  \& {Blouin}}{{Kilic} et~al.}{2020}]{kilic2020}
{Kilic} M.,  {Bergeron} P.,  {Kosakowski} A.,  {Brown} W.~R.,  {Agueros} M.~A.,
    {Blouin} S.,  2020, arXiv e-prints, \href
  {https://ui.adsabs.harvard.edu/abs/2020arXiv200600323K} {p. arXiv:2006.00323}

\bibitem[\protect\citeauthoryear{{Koester}, {Weidemann}  \&
  {Zeidler}}{{Koester} et~al.}{1982}]{koester82}
{Koester} D.,  {Weidemann} V.,   {Zeidler} E.~M.,  1982, \aap, \href
  {https://ui.adsabs.harvard.edu/abs/1982A&A...116..147K} {116, 147}

\bibitem[\protect\citeauthoryear{{Koester}, {Voss}, {Napiwotzki}, {Christlieb},
  {Homeier}, {Lisker}, {Reimers}  \& {Heber}}{{Koester}
  et~al.}{2009}]{koester09}
{Koester} D.,  {Voss} B.,  {Napiwotzki} R.,  {Christlieb} N.,  {Homeier} D.,
  {Lisker} T.,  {Reimers} D.,   {Heber} U.,  2009, \mn@doi [\aap]
  {10.1051/0004-6361/200912531}, \href
  {https://ui.adsabs.harvard.edu/abs/2009A&A...505..441K} {505, 441}

\bibitem[\protect\citeauthoryear{{Kowalski} \& {Saumon}}{{Kowalski} \&
  {Saumon}}{2006}]{kowalski2006}
{Kowalski} P.~M.,  {Saumon} D.,  2006, \mn@doi [\apjl] {10.1086/509723}, \href
  {https://ui.adsabs.harvard.edu/abs/2006ApJ...651L.137K} {651, L137}

\bibitem[\protect\citeauthoryear{{Landstreet} \& {Bagnulo}}{{Landstreet} \&
  {Bagnulo}}{2019}]{landstreet2019}
{Landstreet} J.~D.,  {Bagnulo} S.,  2019, \mn@doi [\aap]
  {10.1051/0004-6361/201936009}, \href
  {https://ui.adsabs.harvard.edu/abs/2019A&A...628A...1L} {628, A1}

\bibitem[\protect\citeauthoryear{{Landstreet} \& {Bagnulo}}{{Landstreet} \&
  {Bagnulo}}{2020}]{landstreet2020}
{Landstreet} J.~D.,  {Bagnulo} S.,  2020, \mn@doi [\aap]
  {10.1051/0004-6361/201937301}, \href
  {https://ui.adsabs.harvard.edu/abs/2020A&A...634L..10L} {634, L10}

\bibitem[\protect\citeauthoryear{{Leggett}, {Ruiz}  \& {Bergeron}}{{Leggett}
  et~al.}{1998}]{leggett98}
{Leggett} S.~K.,  {Ruiz} M.~T.,   {Bergeron} P.,  1998, \mn@doi [\apj]
  {10.1086/305463}, \href
  {https://ui.adsabs.harvard.edu/abs/1998ApJ...497..294L} {497, 294}

\bibitem[\protect\citeauthoryear{{Leggett} et~al.,}{{Leggett}
  et~al.}{2018}]{leggett2018}
{Leggett} S.~K.,  et~al., 2018, \mn@doi [\apjs] {10.3847/1538-4365/aae7ca},
  \href {https://ui.adsabs.harvard.edu/abs/2018ApJS..239...26L} {239, 26}

\bibitem[\protect\citeauthoryear{{Liebert}, {Dahn}  \& {Monet}}{{Liebert}
  et~al.}{1988}]{liebert1988}
{Liebert} J.,  {Dahn} C.~C.,   {Monet} D.~G.,  1988, \mn@doi [\apj]
  {10.1086/166699}, \href
  {https://ui.adsabs.harvard.edu/abs/1988ApJ...332..891L} {332, 891}

\bibitem[\protect\citeauthoryear{{Liebert}, {Bergeron}, {Schmidt}  \&
  {Saffer}}{{Liebert} et~al.}{1993}]{liebert1993}
{Liebert} J.,  {Bergeron} P.,  {Schmidt} G.~D.,   {Saffer} R.~A.,  1993,
  \mn@doi [\apj] {10.1086/173403}, \href
  {https://ui.adsabs.harvard.edu/abs/1993ApJ...418..426L} {418, 426}

\bibitem[\protect\citeauthoryear{{Limoges}, {Bergeron}  \&
  {L{\'e}pine}}{{Limoges} et~al.}{2015}]{limoges2015}
{Limoges} M.~M.,  {Bergeron} P.,   {L{\'e}pine} S.,  2015, \mn@doi [\apjs]
  {10.1088/0067-0049/219/2/19}, \href
  {https://ui.adsabs.harvard.edu/abs/2015ApJS..219...19L} {219, 19}

\bibitem[\protect\citeauthoryear{{MacDonald} \& {Vennes}}{{MacDonald} \&
  {Vennes}}{1991}]{macdonald1991}
{MacDonald} J.,  {Vennes} S.,  1991, \mn@doi [\apj] {10.1086/169937}, \href
  {https://ui.adsabs.harvard.edu/abs/1991ApJ...371..719M} {371, 719}

\bibitem[\protect\citeauthoryear{{Maxted}, {Marsh}, {Moran}  \& {Han}}{{Maxted}
  et~al.}{2000a}]{maxted2000b}
{Maxted} P.~F.~L.,  {Marsh} T.~R.,  {Moran} C.~K.~J.,   {Han} Z.,  2000a,
  \mn@doi [\mnras] {10.1046/j.1365-8711.2000.03343.x}, \href
  {https://ui.adsabs.harvard.edu/abs/2000MNRAS.314..334M} {314, 334}

\bibitem[\protect\citeauthoryear{{Maxted}, {Marsh}  \& {Moran}}{{Maxted}
  et~al.}{2000b}]{maxted2000a}
{Maxted} P.~F.~L.,  {Marsh} T.~R.,   {Moran} C.~K.~J.,  2000b, \mn@doi [\mnras]
  {10.1046/j.1365-8711.2000.03840.x}, \href
  {https://ui.adsabs.harvard.edu/abs/2000MNRAS.319..305M} {319, 305}

\bibitem[\protect\citeauthoryear{{Minchev}, {Chiappini}  \& {Martig}}{{Minchev}
  et~al.}{2013}]{minchev2013}
{Minchev} I.,  {Chiappini} C.,   {Martig} M.,  2013, \mn@doi [\aap]
  {10.1051/0004-6361/201220189}, \href
  {https://ui.adsabs.harvard.edu/abs/2013A&A...558A...9M} {558, A9}

\bibitem[\protect\citeauthoryear{{Morrell} \& {Naylor}}{{Morrell} \&
  {Naylor}}{2019}]{morrell2019}
{Morrell} S.,  {Naylor} T.,  2019, \mn@doi [\mnras] {10.1093/mnras/stz2242},
  \href {https://ui.adsabs.harvard.edu/abs/2019MNRAS.489.2615M} {489, 2615}

\bibitem[\protect\citeauthoryear{{Munn} et~al.,}{{Munn}
  et~al.}{2017}]{munn2017}
{Munn} J.~A.,  et~al., 2017, \mn@doi [\aj] {10.3847/1538-3881/153/1/10}, \href
  {https://ui.adsabs.harvard.edu/abs/2017AJ....153...10M} {153, 10}

\bibitem[\protect\citeauthoryear{{Ourique}, {Romero}, {Kepler}, {Koester}  \&
  {Amaral}}{{Ourique} et~al.}{2019}]{ourique2019}
{Ourique} G.,  {Romero} A.~D.,  {Kepler} S.~O.,  {Koester} D.,   {Amaral}
  L.~A.,  2019, \mn@doi [\mnras] {10.1093/mnras/sty2751}, \href
  {https://ui.adsabs.harvard.edu/abs/2019MNRAS.482..649O} {482, 649}

\bibitem[\protect\citeauthoryear{{Parsons} et~al.,}{{Parsons}
  et~al.}{2010}]{parsons2010}
{Parsons} S.~G.,  et~al., 2010, \mn@doi [\mnras]
  {10.1111/j.1365-2966.2010.17063.x}, \href
  {https://ui.adsabs.harvard.edu/abs/2010MNRAS.407.2362P} {407, 2362}

\bibitem[\protect\citeauthoryear{{Parsons}, {Rebassa-Mansergas}, {Schreiber},
  {G{\"a}nsicke}, {Zorotovic}  \& {Ren}}{{Parsons} et~al.}{2016}]{parsons2016}
{Parsons} S.~G.,  {Rebassa-Mansergas} A.,  {Schreiber} M.~R.,  {G{\"a}nsicke}
  B.~T.,  {Zorotovic} M.,   {Ren} J.~J.,  2016, \mn@doi [\mnras]
  {10.1093/mnras/stw2143}, \href
  {https://ui.adsabs.harvard.edu/abs/2016MNRAS.463.2125P} {463, 2125}

\bibitem[\protect\citeauthoryear{{Parsons} et~al.,}{{Parsons}
  et~al.}{2018}]{parsons2018}
{Parsons} S.~G.,  et~al., 2018, \mn@doi [\mnras] {10.1093/mnras/sty2345}, \href
  {https://ui.adsabs.harvard.edu/abs/2018MNRAS.481.1083P} {481, 1083}

\bibitem[\protect\citeauthoryear{{Putney}}{{Putney}}{1997}]{putney97}
{Putney} A.,  1997, \mn@doi [\apjs] {10.1086/313037}, \href
  {https://ui.adsabs.harvard.edu/abs/1997ApJS..112..527P} {112, 527}

\bibitem[\protect\citeauthoryear{{Rebassa-Mansergas}, {Ren}, {Parsons},
  {G{\"a}nsicke}, {Schreiber}, {Garc{\'\i}a-Berro}, {Liu}  \&
  {Koester}}{{Rebassa-Mansergas} et~al.}{2016}]{rebassa16}
{Rebassa-Mansergas} A.,  {Ren} J.~J.,  {Parsons} S.~G.,  {G{\"a}nsicke} B.~T.,
  {Schreiber} M.~R.,  {Garc{\'\i}a-Berro} E.,  {Liu} X.~W.,   {Koester} D.,
  2016, \mn@doi [\mnras] {10.1093/mnras/stw554}, \href
  {https://ui.adsabs.harvard.edu/abs/2016MNRAS.458.3808R} {458, 3808}

\bibitem[\protect\citeauthoryear{{Rebassa-Mansergas}
  et~al.,}{{Rebassa-Mansergas} et~al.}{2017}]{rebassa17}
{Rebassa-Mansergas} A.,  et~al., 2017, \mn@doi [\mnras]
  {10.1093/mnras/stx2259}, \href
  {https://ui.adsabs.harvard.edu/abs/2017MNRAS.472.4193R} {472, 4193}

\bibitem[\protect\citeauthoryear{{Rolland}, {Bergeron}  \&
  {Fontaine}}{{Rolland} et~al.}{2018}]{rolland2018}
{Rolland} B.,  {Bergeron} P.,   {Fontaine} G.,  2018, \mn@doi [\apj]
  {10.3847/1538-4357/aab713}, \href
  {https://ui.adsabs.harvard.edu/abs/2018ApJ...857...56R} {857, 56}

\bibitem[\protect\citeauthoryear{{Seabroke} \& {Gilmore}}{{Seabroke} \&
  {Gilmore}}{2007}]{seabroke2007}
{Seabroke} G.~M.,  {Gilmore} G.,  2007, \mn@doi [\mnras]
  {10.1111/j.1365-2966.2007.12210.x}, \href
  {https://ui.adsabs.harvard.edu/abs/2007MNRAS.380.1348S} {380, 1348}

\bibitem[\protect\citeauthoryear{{Sion}, {Greenstein}, {Landstreet}, {Liebert},
  {Shipman}  \& {Wegner}}{{Sion} et~al.}{1983}]{sion83}
{Sion} E.~M.,  {Greenstein} J.~L.,  {Landstreet} J.~D.,  {Liebert} J.,
  {Shipman} H.~L.,   {Wegner} G.~A.,  1983, \mn@doi [\apj] {10.1086/161036},
  \href {https://ui.adsabs.harvard.edu/abs/1983ApJ...269..253S} {269, 253}

\bibitem[\protect\citeauthoryear{{Sion}, {Holberg}, {Oswalt}, {McCook}  \&
  {Wasatonic}}{{Sion} et~al.}{2009}]{sion2009white}
{Sion} E.~M.,  {Holberg} J.~B.,  {Oswalt} T.~D.,  {McCook} G.~P.,   {Wasatonic}
  R.,  2009, \mn@doi [\aj] {10.1088/0004-6256/138/6/1681}, \href
  {https://ui.adsabs.harvard.edu/abs/2009AJ....138.1681S} {138, 1681}

\bibitem[\protect\citeauthoryear{{Subasavage} et~al.,}{{Subasavage}
  et~al.}{2017}]{subasavage2017}
{Subasavage} J.~P.,  et~al., 2017, \mn@doi [\aj] {10.3847/1538-3881/aa76e0},
  \href {https://ui.adsabs.harvard.edu/abs/2017AJ....154...32S} {154, 32}

\bibitem[\protect\citeauthoryear{{Taylor}}{{Taylor}}{2005}]{topcat}
{Taylor} M.~B.,  2005, in {Shopbell} P.,  {Britton} M.,   {Ebert} R.,  eds,
  Astronomical Society of the Pacific Conference Series Vol. 347, Astronomical
  Data Analysis Software and Systems XIV. p.~29

\bibitem[\protect\citeauthoryear{{Temmink}, {Toonen}, {Zapartas}, {Justham}  \&
  {G{\"a}nsicke}}{{Temmink} et~al.}{2019}]{temmink2019}
{Temmink} K.~D.,  {Toonen} S.,  {Zapartas} E.,  {Justham} S.,   {G{\"a}nsicke}
  B.~T.,  2019, arXiv e-prints, \href
  {https://ui.adsabs.harvard.edu/abs/2019arXiv191005335T} {p. arXiv:1910.05335}

\bibitem[\protect\citeauthoryear{{Toonen}, {Hollands}, {G{\"a}nsicke}  \&
  {Boekholt}}{{Toonen} et~al.}{2017}]{toonen2017}
{Toonen} S.,  {Hollands} M.,  {G{\"a}nsicke} B.~T.,   {Boekholt} T.,  2017,
  \mn@doi [\aap] {10.1051/0004-6361/201629978}, \href
  {https://ui.adsabs.harvard.edu/abs/2017A&A...602A..16T} {602, A16}

\bibitem[\protect\citeauthoryear{{Torres}, {Cantero}, {Rebassa-Mansergas},
  {Skorobogatov}, {Jim{\'e}nez-Esteban}  \& {Solano}}{{Torres}
  et~al.}{2019}]{torres19}
{Torres} S.,  {Cantero} C.,  {Rebassa-Mansergas} A.,  {Skorobogatov} G.,
  {Jim{\'e}nez-Esteban} F.~M.,   {Solano} E.,  2019, \mn@doi [\mnras]
  {10.1093/mnras/stz814}, \href
  {https://ui.adsabs.harvard.edu/abs/2019MNRAS.485.5573T} {485, 5573}

\bibitem[\protect\citeauthoryear{{Tremblay} \& {Bergeron}}{{Tremblay} \&
  {Bergeron}}{2008}]{tremblay2008}
{Tremblay} P.~E.,  {Bergeron} P.,  2008, \mn@doi [\apj] {10.1086/524134}, \href
  {https://ui.adsabs.harvard.edu/abs/2008ApJ...672.1144T} {672, 1144}

\bibitem[\protect\citeauthoryear{{Tremblay}, {Ludwig}, {Steffen}, {Bergeron}
  \& {Freytag}}{{Tremblay} et~al.}{2011}]{hydrogenmodelpier2011}
{Tremblay} P.~E.,  {Ludwig} H.~G.,  {Steffen} M.,  {Bergeron} P.,   {Freytag}
  B.,  2011, \mn@doi [\aap] {10.1051/0004-6361/201117310}, \href
  {https://ui.adsabs.harvard.edu/abs/2011A&A...531L..19T} {531, L19}

\bibitem[\protect\citeauthoryear{{Tremblay}, {Ludwig}, {Steffen}  \&
  {Freytag}}{{Tremblay} et~al.}{2013}]{tremblay2013}
{Tremblay} P.~E.,  {Ludwig} H.~G.,  {Steffen} M.,   {Freytag} B.,  2013,
  \mn@doi [\aap] {10.1051/0004-6361/201322318}, \href
  {https://ui.adsabs.harvard.edu/abs/2013A&A...559A.104T} {559, A104}

\bibitem[\protect\citeauthoryear{{Tremblay}, {Leggett}, {Lodieu}, {Freytag},
  {Bergeron}, {Kalirai}  \& {Ludwig}}{{Tremblay}
  et~al.}{2014a}]{mixedmodelpier2014}
{Tremblay} P.~E.,  {Leggett} S.~K.,  {Lodieu} N.,  {Freytag} B.,  {Bergeron}
  P.,  {Kalirai} J.~S.,   {Ludwig} H.~G.,  2014a, \mn@doi [\apj]
  {10.1088/0004-637X/788/2/103}, \href
  {https://ui.adsabs.harvard.edu/abs/2014ApJ...788..103T} {788, 103}

\bibitem[\protect\citeauthoryear{{Tremblay}, {Kalirai}, {Soderblom}, {Cignoni}
  \& {Cummings}}{{Tremblay} et~al.}{2014b}]{tremblay2014}
{Tremblay} P.~E.,  {Kalirai} J.~S.,  {Soderblom} D.~R.,  {Cignoni} M.,
  {Cummings} J.,  2014b, \mn@doi [\apj] {10.1088/0004-637X/791/2/92}, \href
  {https://ui.adsabs.harvard.edu/abs/2014ApJ...791...92T} {791, 92}

\bibitem[\protect\citeauthoryear{{Tremblay}, {Fontaine}, {Freytag}, {Steiner},
  {Ludwig}, {Steffen}, {Wedemeyer}  \& {Brassard}}{{Tremblay}
  et~al.}{2015}]{tremblay2015}
{Tremblay} P.~E.,  {Fontaine} G.,  {Freytag} B.,  {Steiner} O.,  {Ludwig}
  H.~G.,  {Steffen} M.,  {Wedemeyer} S.,   {Brassard} P.,  2015, \mn@doi [\apj]
  {10.1088/0004-637X/812/1/19}, \href
  {https://ui.adsabs.harvard.edu/abs/2015ApJ...812...19T} {812, 19}

\bibitem[\protect\citeauthoryear{{Tremblay}, {Cummings}, {Kalirai},
  {G{\"a}nsicke}, {Gentile-Fusillo}  \& {Raddi}}{{Tremblay}
  et~al.}{2016}]{tremblay2016}
{Tremblay} P.~E.,  {Cummings} J.,  {Kalirai} J.~S.,  {G{\"a}nsicke} B.~T.,
  {Gentile-Fusillo} N.,   {Raddi} R.,  2016, \mn@doi [\mnras]
  {10.1093/mnras/stw1447}, \href
  {https://ui.adsabs.harvard.edu/abs/2016MNRAS.461.2100T} {461, 2100}

\bibitem[\protect\citeauthoryear{{Tremblay}, {Cukanovaite}, {Gentile Fusillo},
  {Cunningham}  \& {Hollands}}{{Tremblay} et~al.}{2019a}]{tremblay2019}
{Tremblay} P.~E.,  {Cukanovaite} E.,  {Gentile Fusillo} N.~P.,  {Cunningham}
  T.,   {Hollands} M.~A.,  2019a, \mn@doi [\mnras] {10.1093/mnras/sty3067},
  \href {https://ui.adsabs.harvard.edu/abs/2019MNRAS.482.5222T} {482, 5222}

\bibitem[\protect\citeauthoryear{{Tremblay} et~al.,}{{Tremblay}
  et~al.}{2019b}]{tremblayNature}
{Tremblay} P.-E.,  et~al., 2019b, \mn@doi [\nat] {10.1038/s41586-018-0791-x},
  \href {https://ui.adsabs.harvard.edu/abs/2019Natur.565..202T} {565, 202}

\bibitem[\protect\citeauthoryear{{Tremblay} et~al.,}{{Tremblay}
  et~al.}{2020}]{tremblay2020_WHT}
{Tremblay} P.~E.,  et~al., 2020, arXiv e-prints, \href
  {https://ui.adsabs.harvard.edu/abs/2020arXiv200600965T} {p. arXiv:2006.00965}

\bibitem[\protect\citeauthoryear{{Veras}, {Tremblay}, {Hermes}, {McDonald},
  {Kennedy}, {Meru}  \& {G{\"a}nsicke}}{{Veras} et~al.}{2020}]{veras2020}
{Veras} D.,  {Tremblay} P.-E.,  {Hermes} J.~J.,  {McDonald} C.~H.,  {Kennedy}
  G.~M.,  {Meru} F.,   {G{\"a}nsicke} B.~T.,  2020, \mn@doi [\mnras]
  {10.1093/mnras/staa241}, \href
  {https://ui.adsabs.harvard.edu/abs/2020MNRAS.493..765V} {493, 765}

\bibitem[\protect\citeauthoryear{{Winters} et~al.,}{{Winters}
  et~al.}{2020}]{winters2020}
{Winters} J.~G.,  et~al., 2020, arXiv e-prints, \href
  {https://ui.adsabs.harvard.edu/abs/2020arXiv200411225W} {p. arXiv:2004.11225}

\bibitem[\protect\citeauthoryear{{Zuckerman}, {Becklin}, {Macintosh}  \&
  {Bida}}{{Zuckerman} et~al.}{1997}]{zuckerman97}
{Zuckerman} B.,  {Becklin} E.~E.,  {Macintosh} B.~A.,   {Bida} T.,  1997,
  \mn@doi [\aj] {10.1086/118296}, \href
  {https://ui.adsabs.harvard.edu/abs/1997AJ....113..764Z} {113, 764}

\bibitem[\protect\citeauthoryear{{Zuckerman}, {Koester}, {Reid}  \&
  {H{\"u}nsch}}{{Zuckerman} et~al.}{2003}]{zuckerman2003}
{Zuckerman} B.,  {Koester} D.,  {Reid} I.~N.,   {H{\"u}nsch} M.,  2003, \mn@doi
  [\apj] {10.1086/377492}, \href
  {https://ui.adsabs.harvard.edu/abs/2003ApJ...596..477Z} {596, 477}

\bibitem[\protect\citeauthoryear{{Zuckerman}, {Melis}, {Klein}, {Koester}  \&
  {Jura}}{{Zuckerman} et~al.}{2010}]{zuckerman2010}
{Zuckerman} B.,  {Melis} C.,  {Klein} B.,  {Koester} D.,   {Jura} M.,  2010,
  \mn@doi [\apj] {10.1088/0004-637X/722/1/725}, \href
  {https://ui.adsabs.harvard.edu/abs/2010ApJ...722..725Z} {722, 725}

\bibitem[\protect\citeauthoryear{{van Altena}, {Lee}  \& {Hoffleit}}{{van
  Altena} et~al.}{1995}]{vanAltena95}
{van Altena} W.~F.,  {Lee} J.~T.,   {Hoffleit} E.~D.,  1995, {The general
  catalogue of trigonometric [stellar] parallaxes}.
New Haven, CT: Yale University Observatory

\bibitem[\protect\citeauthoryear{{van Leeuwen}}{{van
  Leeuwen}}{2007}]{vanLeeuwen07}
{van Leeuwen} F.,  2007, \mn@doi [\aap] {10.1051/0004-6361:20078357}, \href
  {http://adsabs.harvard.edu/abs/2007A\%26A...474..653V} {474, 653}

\makeatother
\end{thebibliography}
